%% file: ms.tex
\documentclass[12pt,preprint]{aastex}
\usepackage{subfigure}
\usepackage{longtable}

\newcommand{\beq}{\begin{equation}}
\newcommand{\eeq}{\end{equation}}

\def\ln{{\rm ln}}

\def\3he{$^3$He\,}
\def\he4{$^4$He\,}

\shorttitle{Exponential Growth}
\shortauthors{Han \& Liu}

\begin{document}

\title{Exponential Growth of the Emission Measure in the Impulsive Phase Derived from X-ray Observations of Solar Flares}

\author{Feiran Han\altaffilmark{1}, and
Siming Liu\altaffilmark{1}
}
\altaffiltext{1}{Key Laboratory of Dark Matter and Space Astronomy, Purple Mountain Observatory, Chinese Academy of Sciences, Nanjing, 210008, P. R. China}

\begin{abstract}

Light curves of solar flares in the impulsive phase are complex in general, which is expected given
complexities of the flare environment in the magnetic field dominant corona. With GOES observations, we however find
that there are a subset of flares, whose impulsive phases are dominated by a period of exponential
growth of the emission measure.
Flares occurring from Jan. 1999 to Dec. 2002 are analyzed, and results from observations made with both GOES 8 and 10 satellites are compared to estimate instrumental uncertainties.
The frequency distribution of the mean temperature during this exponential growth phase has a normal distribution.
Most flares within the 1$\sigma$ range of this temperature distribution belong to GOES class B or C with the frequency distribution of the peak flux of the GOES low-energy channel following a log-normal distribution. The frequency distribution of the growth rate and the duration of the exponential growth phase also follow a log-normal distribution with the duration covering a range from half a minute to about half an hour. As expected, the growth time is correlated with the decay time of the soft X-ray flux. We also find that the growth rate
of the emission measure is strongly anti-correlated with the duration of the exponential growth phase and increases with the mean temperature.
The implications of these results on the study of energy release in solar flares are discussed at the end.

\end{abstract}


\section{INTRODUCTION}
\label{intro}
Most observational studies of solar flares focus on detailed analysis of individual events \citep[e.g.,][]{m94, l03, h08, r09, l10, b11}. Statistical studies usually treat the flare population as a whole to extract a few quantities and analyze their occurrence frequency distribution \citep[e.g.,][]{a94, v02a, v02b, s06}. Although these two approaches are complementary to each other, relations between results of these two studies remain obscure. For example, it is not clear how specific physical processes revealed from studies of individual flares may lead to power-law distributions of the flare occurrence rate with respect to many of their observed characteristics. It is possible that statistical properties of flares are mostly determined by the environment and may not depend on the detailed physical processes \citep{l91}. However, even for flares with similar peak value of the soft X-ray (SXR) flux, an important quantity characterizing the flare amplitude, their appearance can be drastically different, which reflects the intrinsic complexity of flares as a macroscopic phenomena with an enormous degree of freedom. The dominant physical process in one flare may be distinct from that of the other. To better understand the flare phenomena, it is therefore necessary to classify flares based on some of the observed prominent characteristics and explore their physical origin accordingly. A classification of large-scale coronal EIT waves has recently resolved some related controversies \citep{w11}.

One of the key aspects of solar flare study is to explore the energy release processes, and it is generally accepted that the impulsive phase dominates the overall energy release \citep{h11}. However, in contrast to the relatively smooth decay of X-ray fluxes in the gradual phase of most flares, which is associated with coronal loops and has been studied extensively and better understood \citep{r78, a78, s91, c95, k08}, the X-ray light curves in the impulsive phase are generally complex and there appears to be a variety of physical processes involved. Many of the observed complexities of flares originate from the complex magnetic field structure carrying the flaring plasma. To better understand the basic physical processes related to the energy release, one may focus on studying flares with relatively simple structure, in particular those associated with single loops.

\citet{r09} carried out a detailed analysis of a flaring loop, and we also notice that the impulsive phase of this flare is dominated by a period of exponential growth in both the SXR fluxes and the derived emission measure (EM) (Left panel of Fig. \ref{lc}). These relatively simple behaviors of flaring loops may reflect some elementary processes in the flare energy release \citep{g04, lhf10}. Motivated by this observation, in this paper we analyze GOES observations from Jan. 1999 to Dec. 2002 to identify flares with the impulsive phase dominated by a period of exponential growth of the EM. These flares are a subset of flares. Detailed studies of them may help to reveal the physics of energy release in the impulsive phase in general.

In \S\ \ref{data}, we present the analysis of GOES data and the flare selection criteria. The results are shown in \S\ \ref{res}. These results are discussed in \S\ \ref{con}, where we also draw the conclusions.

\section{Data Analysis}
\label{data}

{\bf Background Selection and Peak Time:} Both GOES 8 and 10 satellites cover the previous maximum of solar activity. We focus on a 4 year period of the activity peak from Jan. 1999 to Dec. 2002. RHESSI was launched into orbit in Feb. 2002. Some of these flares were observed by RHESSI as well \citep{l03}. To derive the temperature and EM of the flaring plasma with the GOES data, it is essential to subtract the pre-flare background fluxes properly \citep{b90}. GOES satellites measure SXR fluxes from the Sun in two wave bands --- 1-8 \AA\ and 0.5-4 \AA\ --- with a cadence of 3 seconds \citep{g94}. We make use of the GOES flare list from http://umbra.nascom.nasa.gov/sdb/ngdc/xray\_events/, where the flare onset time, the peak time of the flux in the lower energy channel, and the end time are given for 10511 flares. The difference between the former two may be called the duration of the flare rise phase: $t_r$. We extract data for these flares from GOES 8 and 10 observations.  For each flare, we extend the range of data analysis both before the flare onset time and after the flare end time by $t_r$ \footnote{For the flare studied by \citet{r09}, the onset time in the flare list is way behind the actual flare onset. For this flare, the range of data analysis is extended by $2t_r$ before the listed flare onset time and after the flare end time.}. For some flares, the peak time in the flare list does not correspond to the maximum of the flux in the low energy channel between the onset and end time. We redefine the flare peak time as the time when the flux reaches its maximum value between the onset and end time. The background fluxes are chosen as that of a period before the flare peak time with a relatively low and constant flux level and are selected independently for the two energy channels \footnote{In practice, we fit the light curve with a set of line segments. The error is assumed to be the same as the measured flux and the critical value of the $\chi^2$ is set at 0.001. The fit starts from the first three data points and the corresponding $\chi^2$ is calculated. If the $\chi^2$ is less than the critical value, we include one more data point following this period for a new linear fit. This process is repeated. A new segment starts whenever the $\chi^2$ of the current segment reaches this critical value. We then calculate the mean value of the flux for each line segment. For the two segments with the lowest mean fluxes before the flare peak time, we set the flux of the segment with a lower gradient as the background flux.}. In the following, we will mostly use results from the low-energy channel, where the background flux is high and the signal relatively weak, to define the flare characteristics.

\begin{figure}[ht]
\centering
\subfigure[]{ \label{bk1}
\includegraphics[width=8cm]{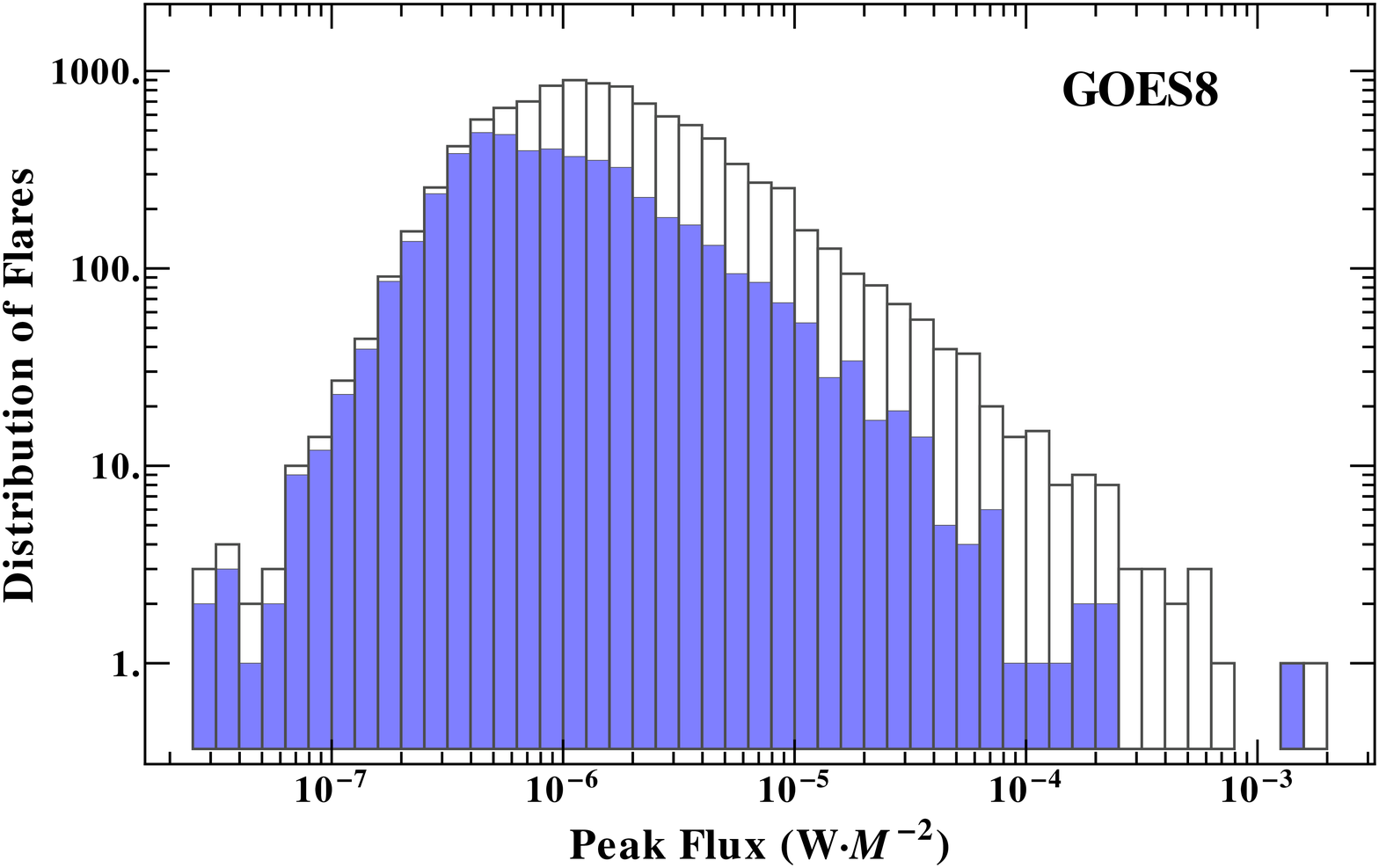}}
\subfigure[]{ \label{bk2}
\includegraphics[width=8cm]{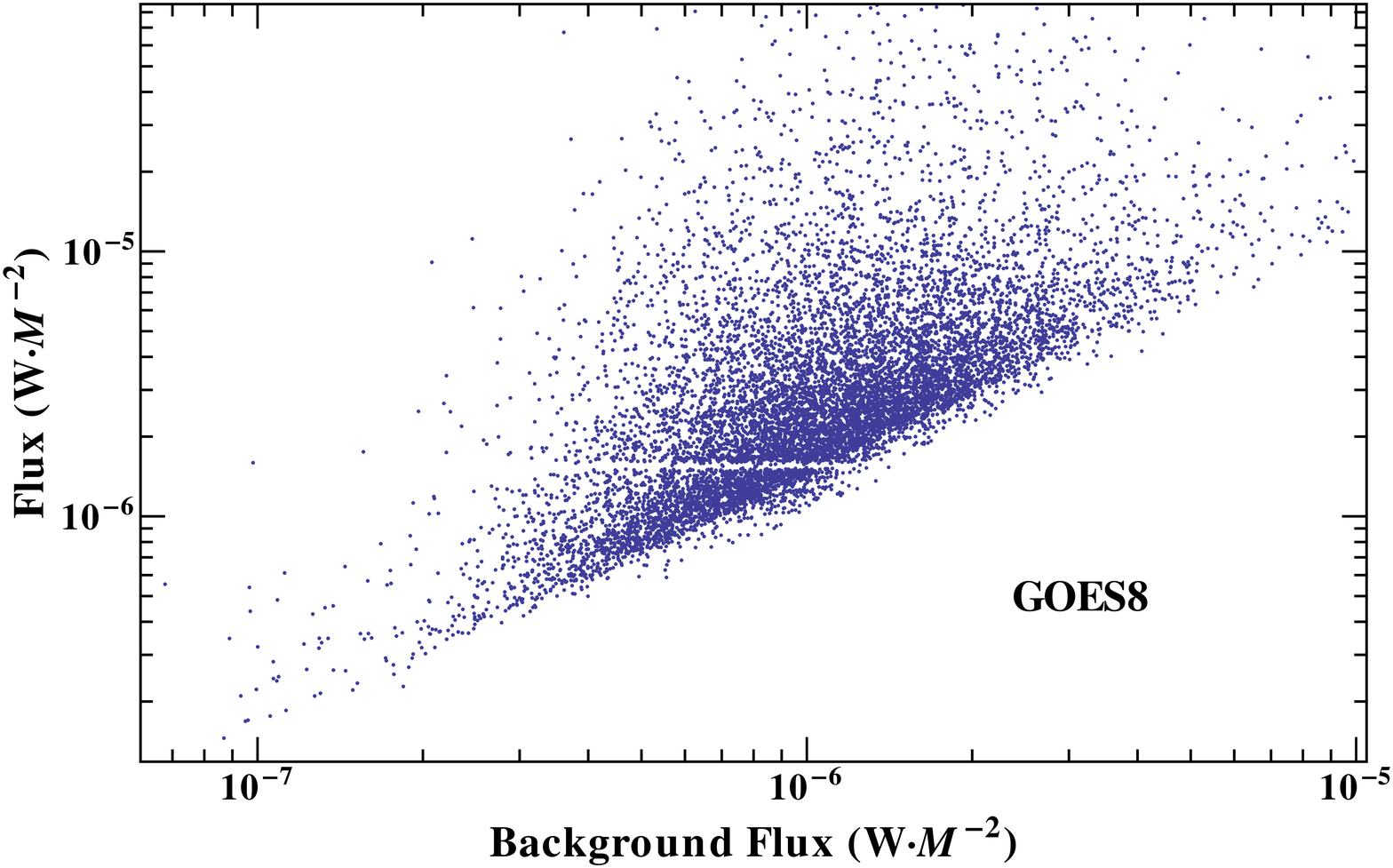}}
\subfigure[]{ \label{bk3}
\includegraphics[width=8cm]{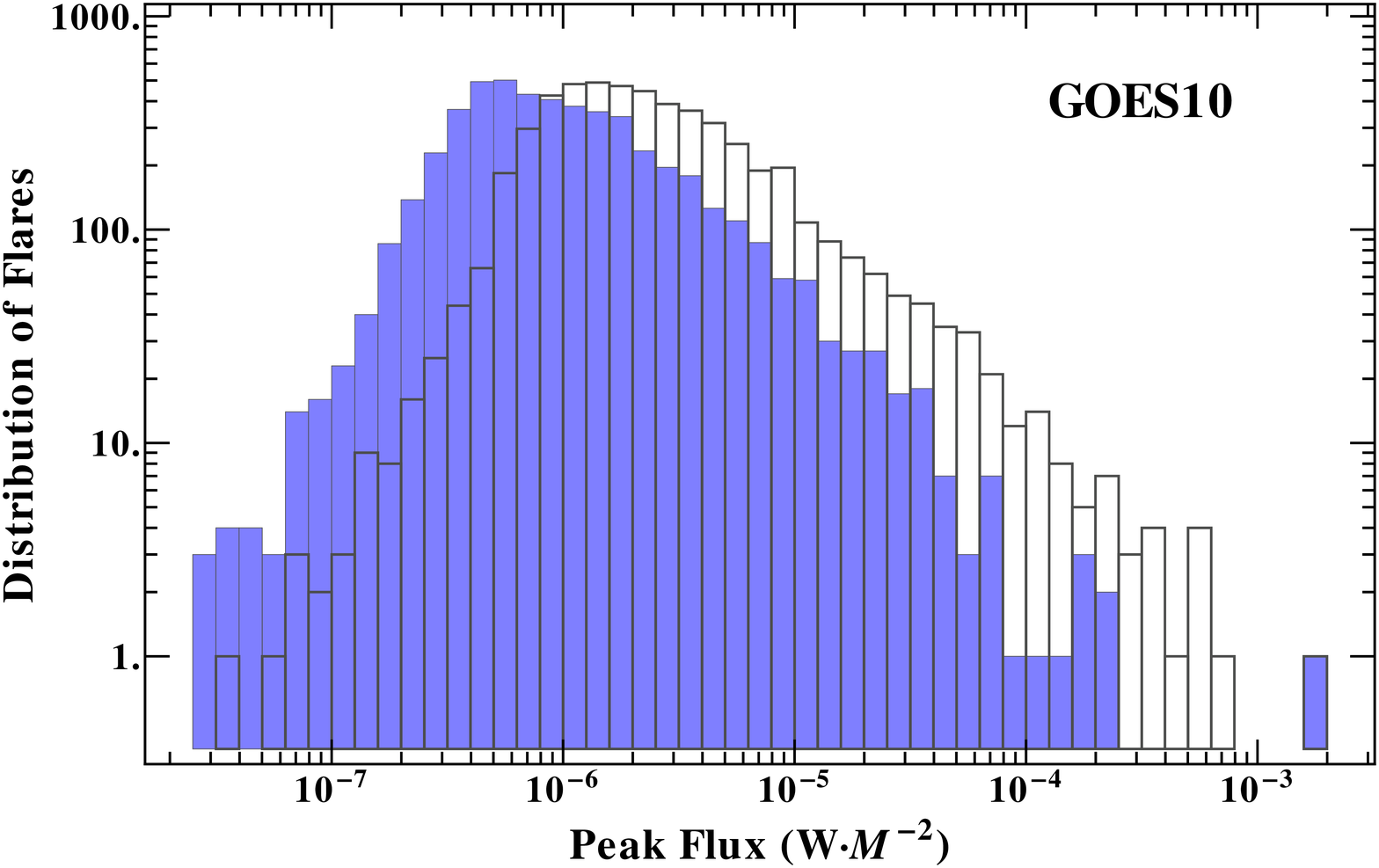}}
\subfigure[]{ \label{bk4}
\includegraphics[width=8cm]{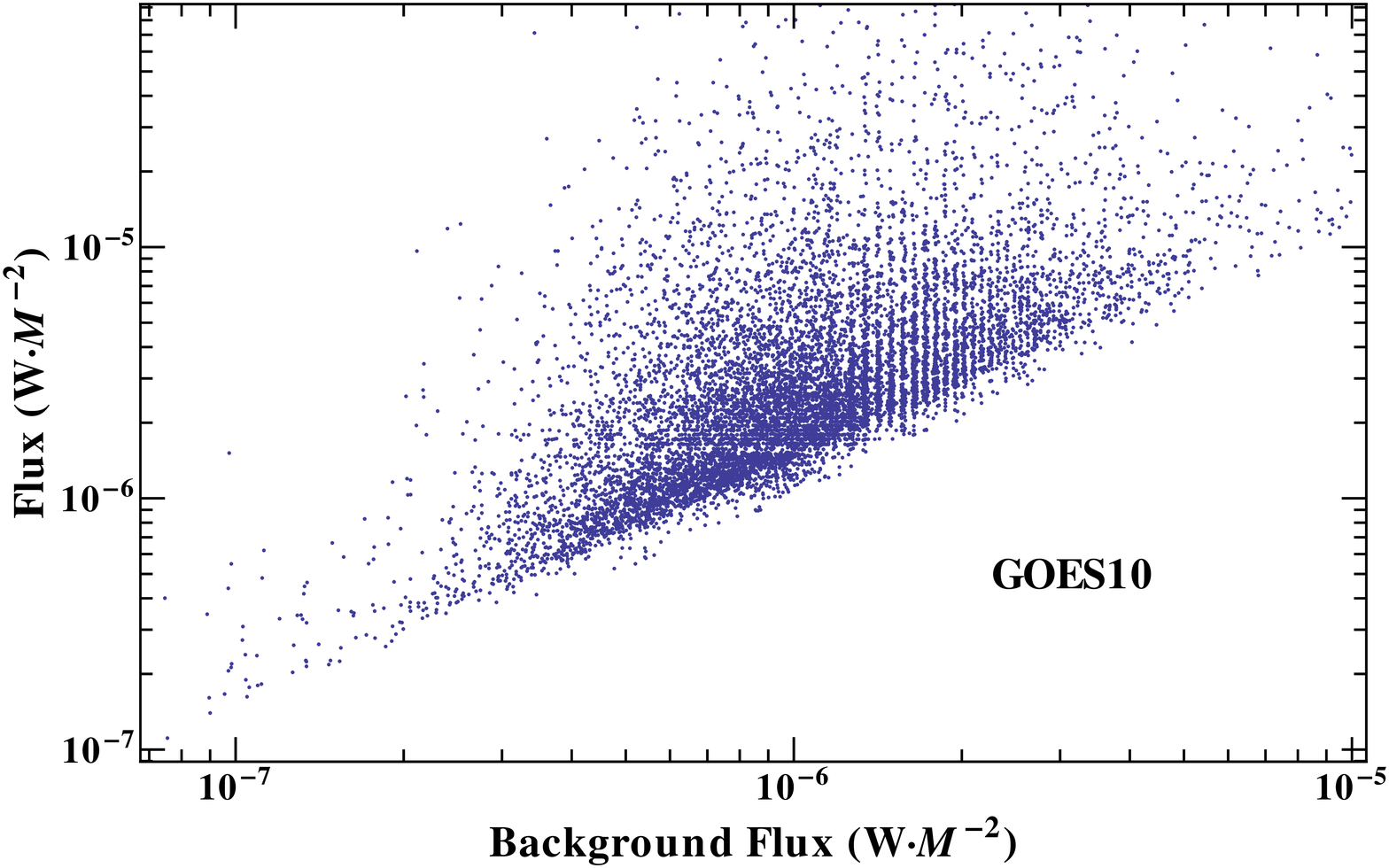}}
\caption{Left: Background subtracted peak flux distribution. The shaded histogram is for flares with a background flux less than or equal to $10^{-6}$W M$^{-2}$. The other histogram is for the rest of flares with a higher pre-flare background flux. Right: Correlation between the background flux and the peak flux (including the pre-flare background flux).
}
\label{bk}
\end{figure}
The left panels of Figure \ref{bk} show the frequency distribution of the background subtracted peak flux of all flares. The shaded histograms are for flares with a low level of pre-flare background flux ($\leq10^{-6}$W M$^{-2}$). The other histograms are for all other flares with a high level of pre-flare background. The distributions of flares with low and high pre-flare background flux agree with each other at high peak fluxes, which is consistent with the prediction of the self-organized criticality model of \citet{l91}. The instrumental bias is important at low peak fluxes, and the difference between GOES 8 and 10 observations is obvious.

The right panels of Figure \ref{bk} show the correlation between the pre-flare background flux and the flux at the peak time. For a flare to be identified in the data, the one-minute averaged peak flux needs to exceed the pre-flare background flux by at least 40\%, which explains why there are no flares in the low-right side\footnote{http://umbra.nascom.nasa.gov/sdb/ngdc/xray\_events/xraytime.txt}. The seeming correlation between these two fluxes therefore is mostly caused by this flare identification procedure. The distribution in the left panels show that the occurrence chance of flares of a given amplitude does not depend on the pre-flare background flux. Actually the pre-flare background flux is mostly caused by decay of earlier flares \citep{a94}. These results show that most flares studied here are independent from each other.
The obvious horizontal strip for GOES 8 and horizontal and vertical strips for GOES 10 observations are caused by the digitalization process of the instrument \citep{g94}.

\begin{figure}[!ht]
\centering
\includegraphics[width=8cm,height=13cm]{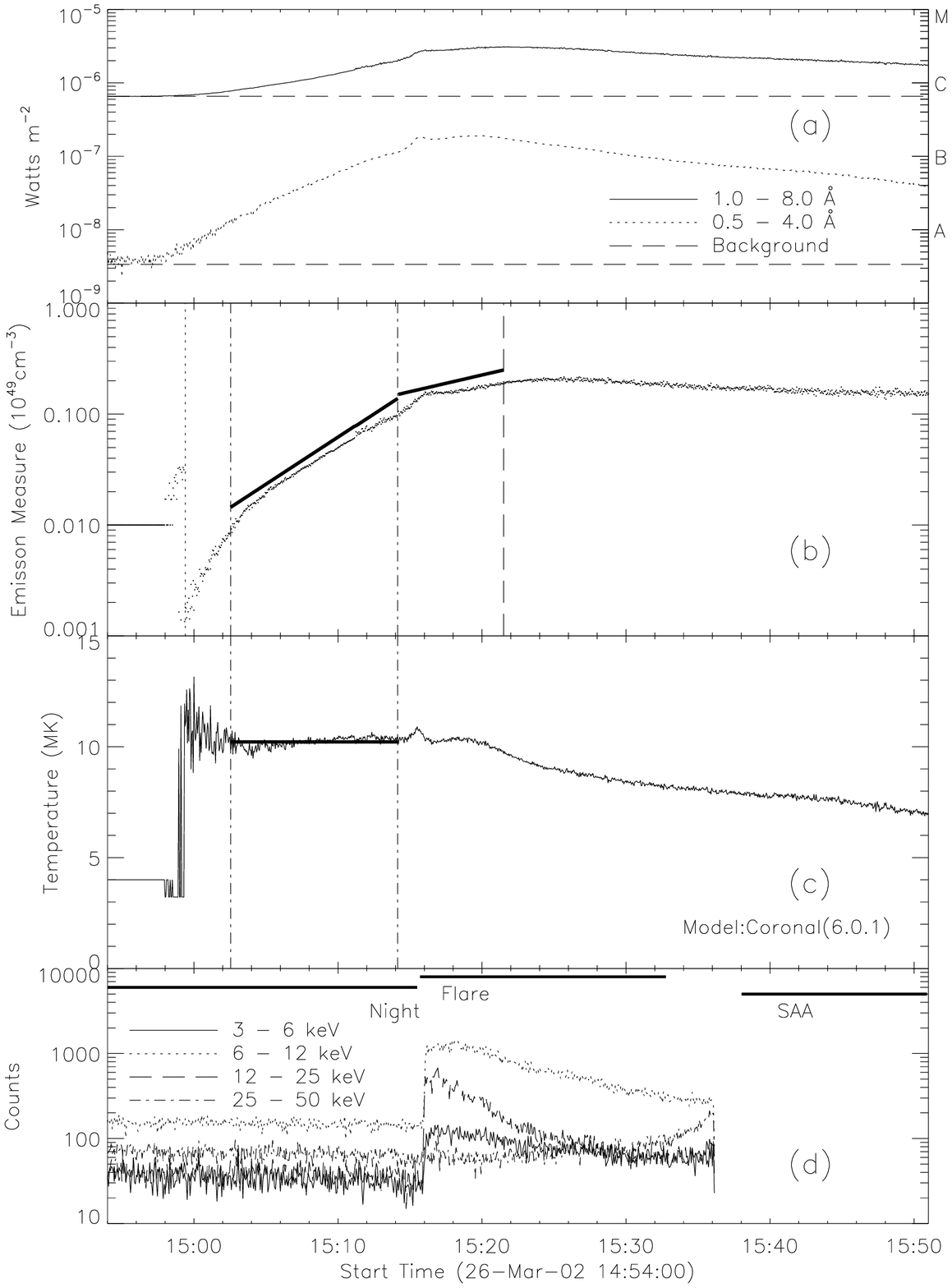}
\hfill
\includegraphics[width=8cm,height=13cm]{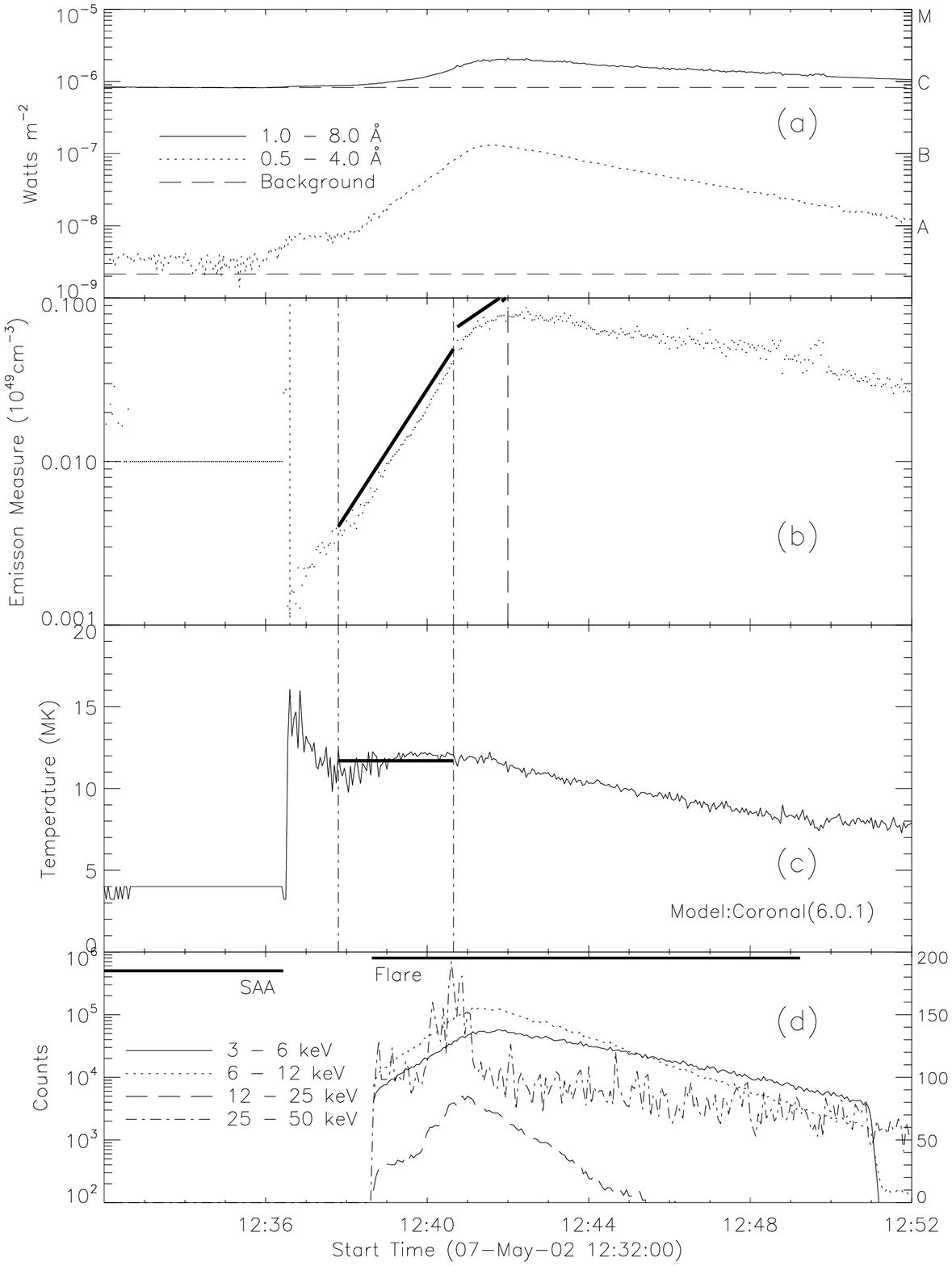}
\caption{Observation summary of two flares on 2002 March 26 (left) and 2002 May 7 (right). (a) GOES 8 light curves (3 s data). The dashed lines indicate the background fluxes.
(b) EM derived from the GOES fluxes. Solid line segments illustrate the periods of exponential growth in EM. The dominant exponential growth phase of the EM is indicated by the two vertical dot-dashed lines. The vertical dotted and dashed lines indicate the flare onset and peak times, respectively. (c) Temperature derived from the GOES fluxes. The mean temperature during the dominant exponential growth phase is indicated by a solid horizontal line. (d) RHESSI light curves (4 s data) in different energy bands. RHESSI night and the South Atlantic Anomaly (SAA) time intervals are indicated by horizontal lines. In the right panel, RHESSI light curve in 25-50 keV energy band is plotted in linear scale to show its impulsive behavior.}
\label{lc}
\end{figure}

{\bf Onset Time:} With the background fluxes selected, we redefine the onset time to better quantify the rise phase. The background subtracted flux (in the low energy channel) needs to exceed $2.1\times 10^{-8}$W m$^{-2}$ to obtain reliable temperature and EM \citep{g94}.  For a background subtract flux below this critical value, the GOES software gives a default value of 4 MK and $0.01\times 10^{49}$ cm$^{-3}$ for the temperature and EM, respectively (See Fig. \ref{lc})\footnote{See SSWIDL program: /ssw/gen/idl/synoptic/goes/goes\_chianti\_tem.pro}. The background subtracted flux therefore needs to exceed this critical value after the flare onset. Similarly, we require that the background subtracted flux in the high energy channel should be greater than $1.0\times 10^{-10}$W m$^{-2}$ \footnote{See SSWIDL program: /ssw/gen/idl/synoptic/goes/goes\_tem\_calc.pro}.
We use the Coronal emission model version 6.0.1 to derive the temperature and EM. In the early rise phase, the signal may be weak so that the obtained EM and temperature can fluctuate significantly. We require that after the flare onset, the difference of the logarithm of the EM between two neighboring data points should not exceed $15\%$ of the difference between the maximum and minimum values of the logarithm of the EM of the flare data range.

{\bf Segments of Exponential Growth of the EM:}  We are mostly interested in the rise phase. To identify periods of exponential growth, we fit the time variation of the logarithm of the EM with a set of line segments. Specifically, starting from the peak time of the SXR flux, we do a linear fit to the logarithm of the EM of a period ending at the peak time and calculate the corresponding reduced $\chi^2$. The error has been assumed to be 1. We adjust the duration of this period until the reduced $\chi^2$ reaches a value just below a prior chosen critical value, which gives a segment of approximately exponential growth phase right before the peak time of the SXR flux. Following a similar procedure, we identify the next segment of exponential growth before the first segment and all other segments before the peak time. In this study, this critical value of the reduced $\chi^2$ is taken as $9.3\times 10^{-4}$, and we exclude flares, whose rise phase can be fitted with a single line segment. Such flares are usually weak and have a short rise phase, the corresponding signals are not reliable.

\begin{figure}[ht]
\centering
\subfigure[]{ \label{srn.sub.1}
\includegraphics[width=8cm]{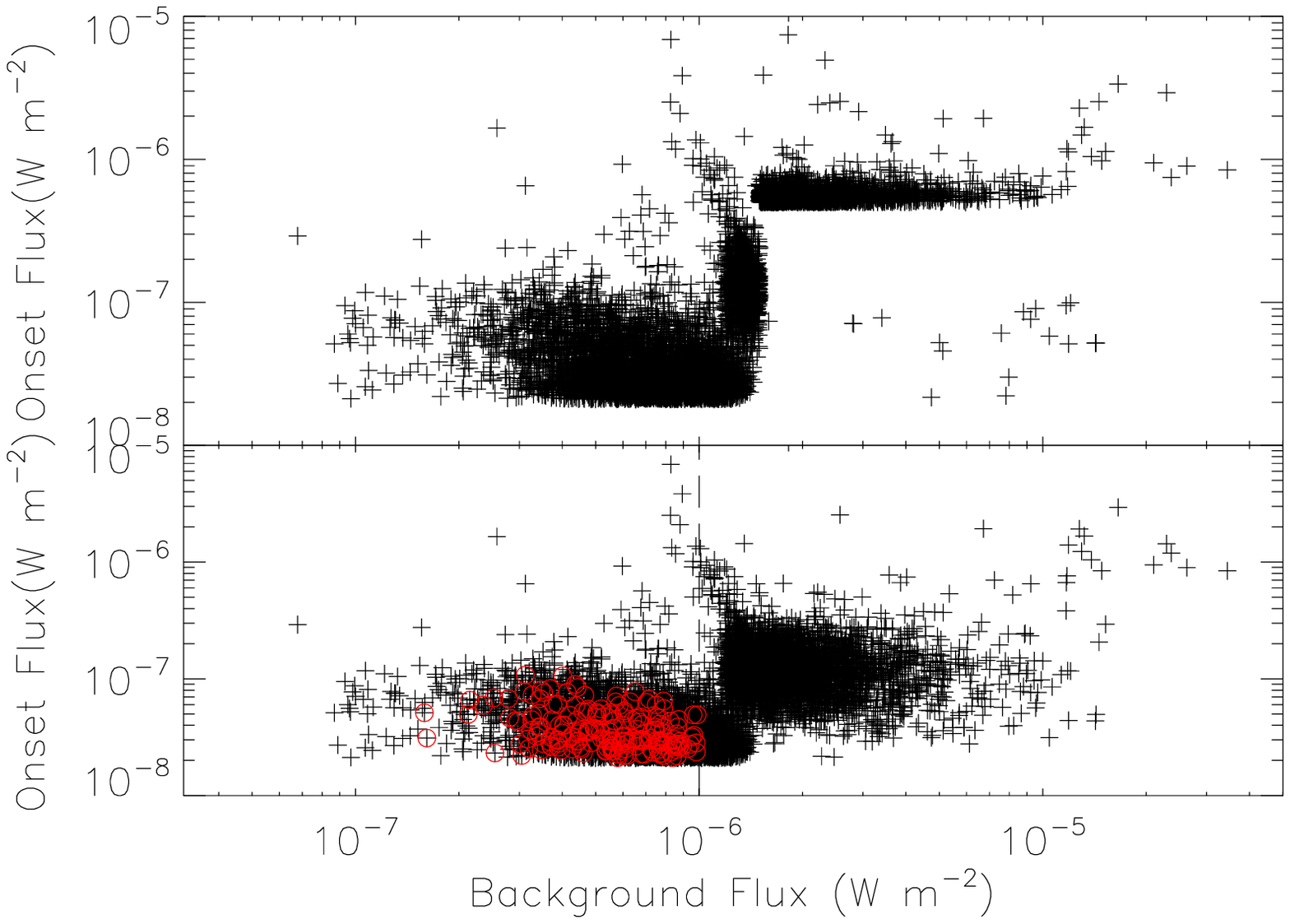}}
\subfigure[]{ \label{srn.sub.2}
\includegraphics[width=8cm]{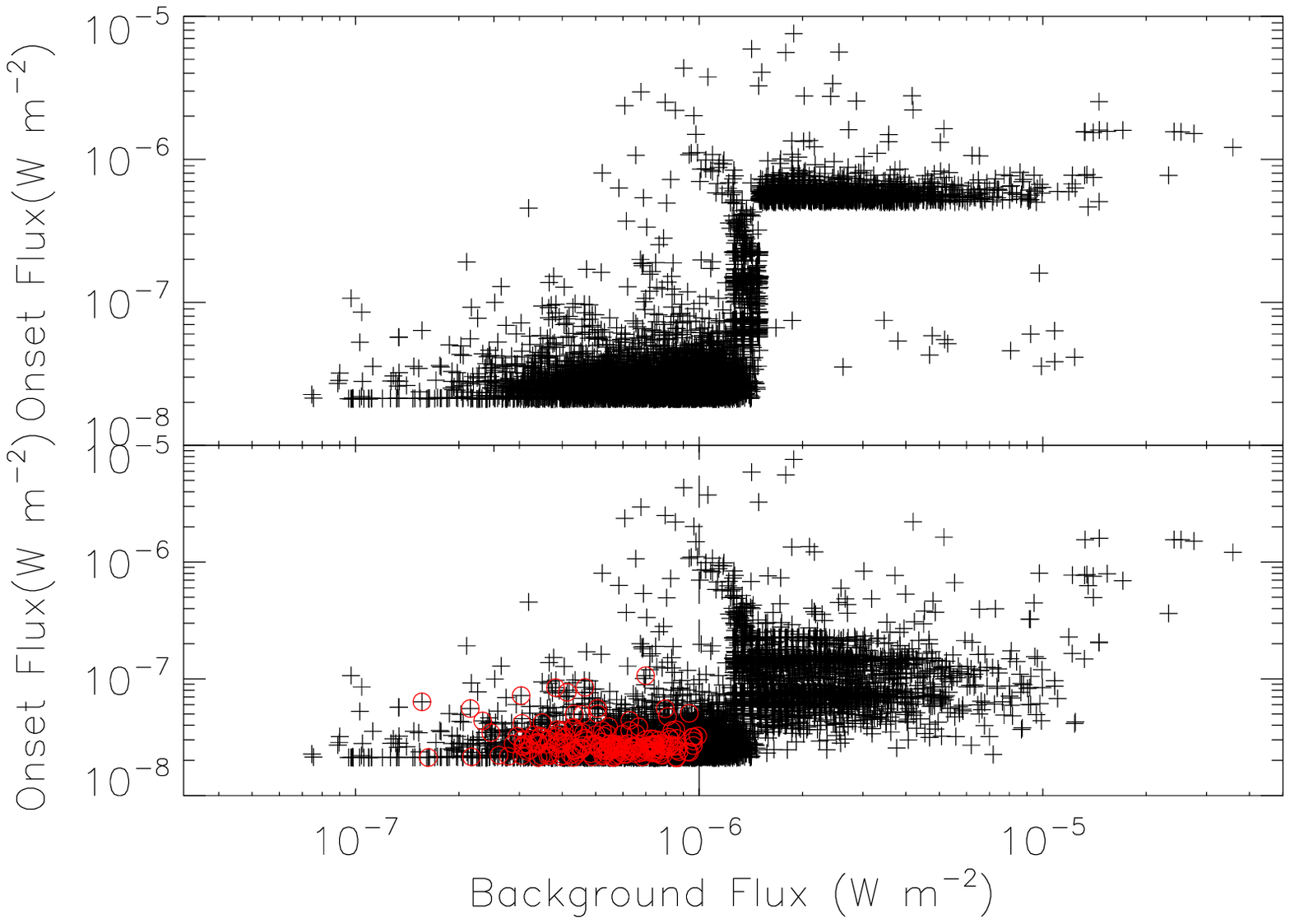}}
\caption{The background flux vs the background subtracted flux at the flare onset in the 1-8 \AA\ energy band of all flares. The left and right panels are for GOES 8 and 10 observations, respectively. The upper panels are obtained with the default GOES software. The lower panels are obtained by removing a criterium on the background flux. The vertical dashed line indicates the upper limit of the background flux for our flare selection. The open circles indicate the selected flares with the rise phase dominated by an exponential growth period of the EM.
}
\label{srn.sub}
\end{figure}

{\bf Instrumental and Software Effects:} With the background flux and the flare onset time determined, we can show all flares on the parameter space of the background flux and the background subtracted flux at the flare onset. Figures \ref{srn.sub.1} and \ref{srn.sub.2} show the results for observations with GOES 8 and 10 satellites, respectively. The upper panels show the results with the GOES default software. There are 10198 and 10230 flares for GOES 8 and 10 observations, respectively. Flares with poor signals and a few outliers outside the range of parameter space shown in these figures have been excluded. It is evident that there are some artifacts caused by either instrumental or software effects. When the background flux exceeds $1.5\times 10^{-6}$ W m$^{-2}$, the default GOES software sets a high threshold of $5\times 10^{-7}$ W m$^{-2}$ for calculation of the temperature and EM \footnote{See SSWIDL program/ssw/gen/idl/synoptic/goes/goes\_tem.pro}, which causes the sharp cut at these background and background subtracted flux levels. Not surprisingly there is a sharp cut at the critical value of $2.1\times 10^{-8}$ W m$^{-2}$, below which the temperature and EM are set to the default values. There are a few outliers, some of which correspond to data gaps. The few outliers in the low-right corner correspond to flares, whose difference in the logarithm of the EM between two neighboring points are always less than $15\%$ of the difference between the maximum and minimum values of that of the flare data range. The flare onset is therefore triggered by the requirement that the background subtracted flux exceeds the critical value of
$2.1\times 10^{-8}$ W m$^{-2}$.

\begin{figure}[ht]
\centering
\subfigure[]{ \label{srnh.sub.1}
\includegraphics[width=8cm]{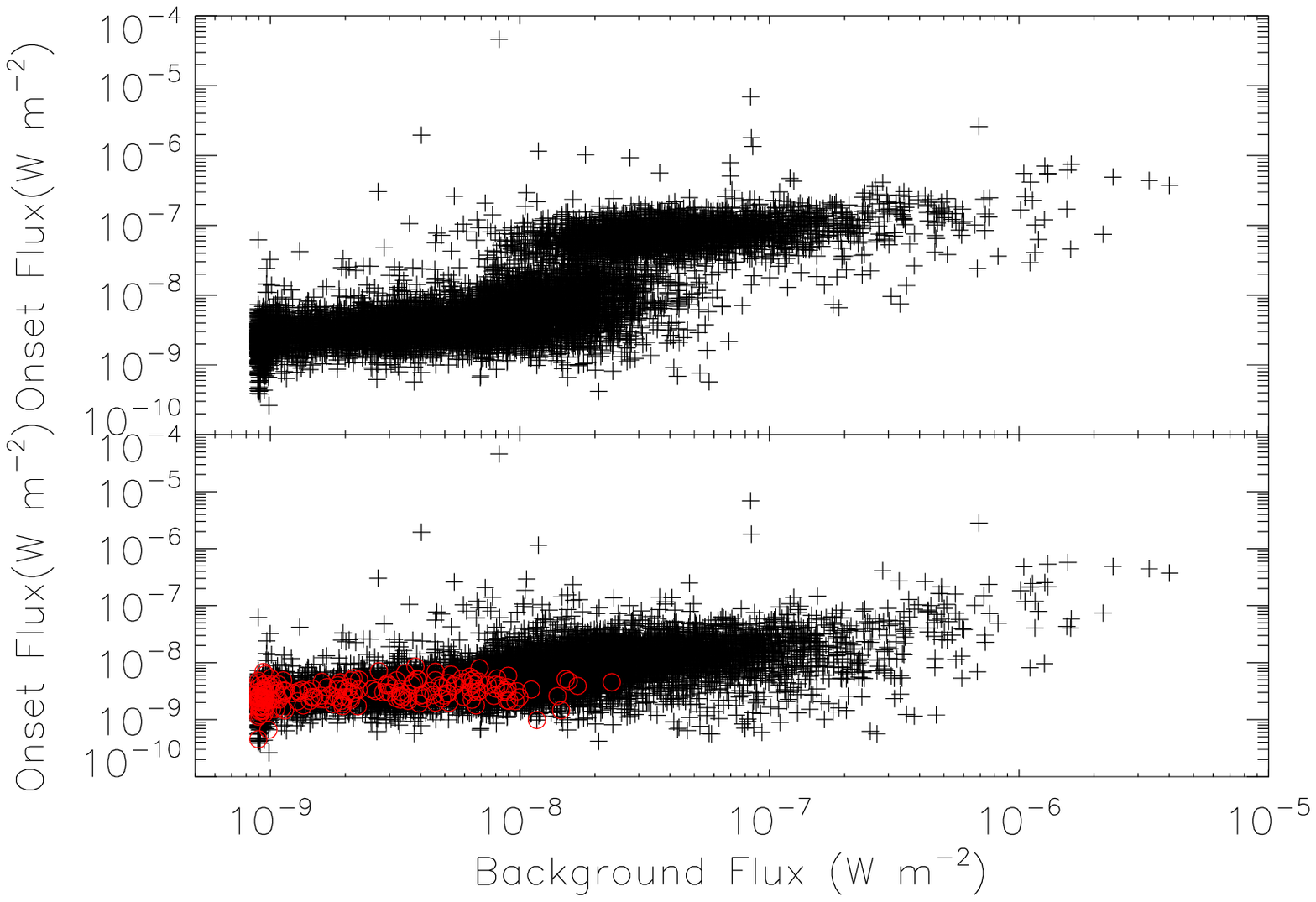}}
\subfigure[]{ \label{srnh.sub.2}
\includegraphics[width=8cm]{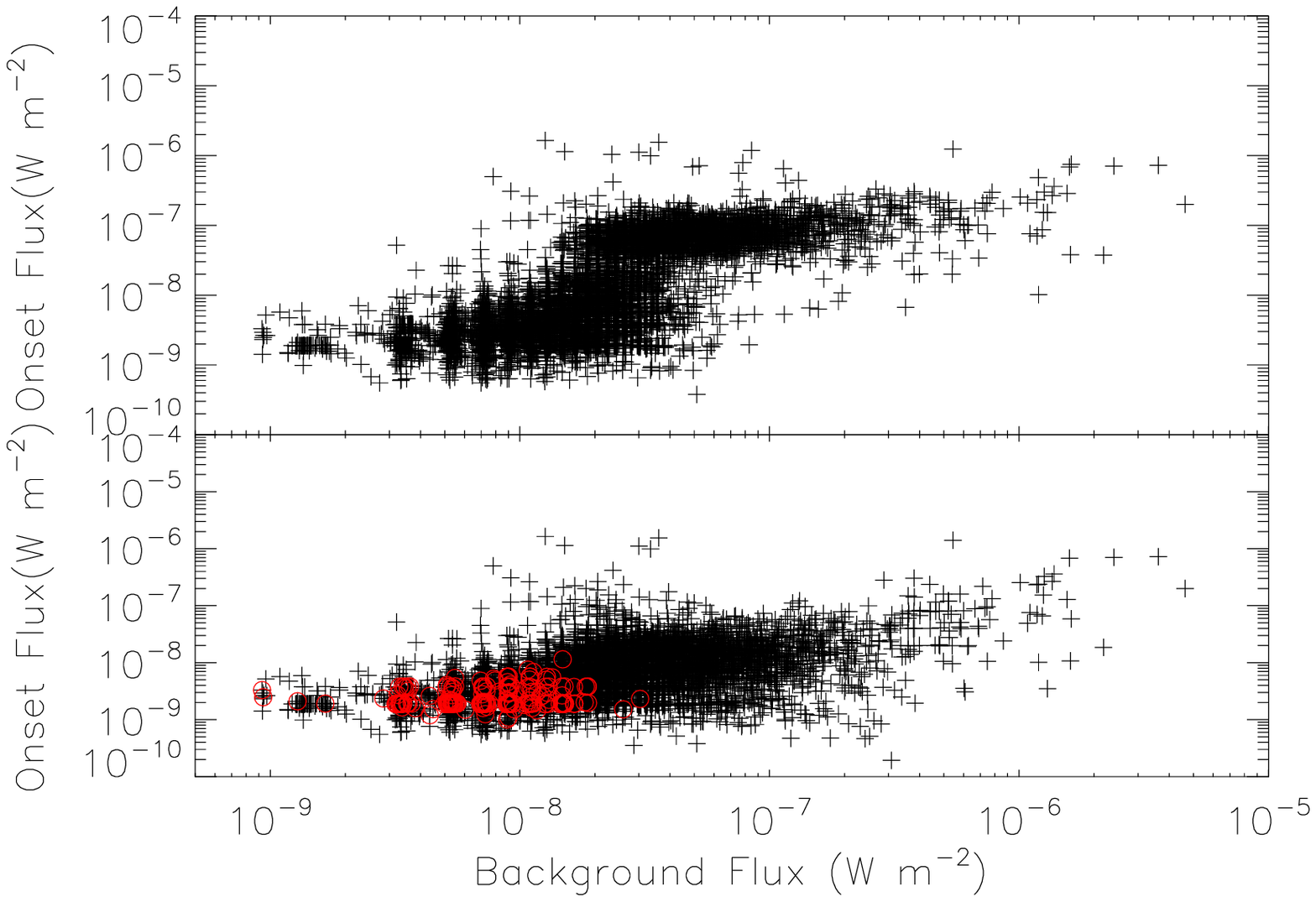}}
\caption{Same as Figure \ref{srn.sub} but for the high energy channel of 0.5-4 \AA.\ Note that the apparent two groups of the distribution in the upper panels are caused by the requirement of the low energy channel flux exceeding the critical value of $5\times10^{-7}$W m$^{-2}$ to give reliable values of the EM and temperature of the default GOES software. These two groups merge in the lower panels.
}
\label{srnh.sub}
\end{figure}

The lower panels are obtained by removing the threshold at $5\times 10^{-7}$ W m$^{-2}$ in the software. There are 10251 and 10275 flares from the GOES 8 and 10 observations, respectively. It is evident that after removing some artifact caused by the software, the instruments have at least two prominent states of response in the low energy channel, which are divided roughly by the background flux level. Moreover, when the background flux is high, the GOES 10 observations show that flares distribute in a few strips, which is likely caused by the digitalization process. A similar plot for the high energy channel, however, shows a more or less continuous distribution (See lower panels of Fig. \ref{srnh.sub}). The two states of response therefore only exist for the low energy channel.

{\bf Flare Selection:} We identify simple flares, whose rise phase is dominated by an exponential growth segment of the
EM with the following criteria: 1) Since the duration of the exponential growth phase is an important quantity to extract, we focus on flares with the background flux in the low energy channel not exceeding $1\times10^{-6}$ W m$^{-2}$, which is indicated by the dashed line in the lower panels of Figure \ref{srn.sub}. When the background flux is high, the signal in the early rise phase may be too low to give reliable temperature and EM measurement so that the observed duration of the dominant segment of exponential growth in the EM becomes shorter. This constraint also addresses the instrumental effects discussed above.
2) The duration of the longest line segment of the logarithm of the EM must exceed 30 seconds and longer than the half length of the rise phase from the flare onset time to the flare peak time. The former criterium ensures that the period of exponential growth is prominent, and the latter ensures its dominance in the rise phase.
3) The increase of the logarithm of the EM during this line segment of exponential growth must exceed 40\% of the difference of the maximum to minimum value of the logarithm of the EM during the rise phase. These two criteria define the dominance of the exponential growth phase.
4) To ensure the simplicity of the selected flares, we also have a linear fit (with an error of 1) to the logarithm of the background subtracted flux in the low-energy channel between the flare peak time and the time when the flux decreases to one half of the peak flux in the decay phase. The reduced $\chi^2$ of this linear fit must be less than $10^{-4}$ for a flare to be selected. Flares with higher values of the reduced $\chi^2$ have more complicated decay phase and are likely associated with multiple loops.

\begin{figure}[!ht]
\centering
\includegraphics[width=8cm,height=13cm]{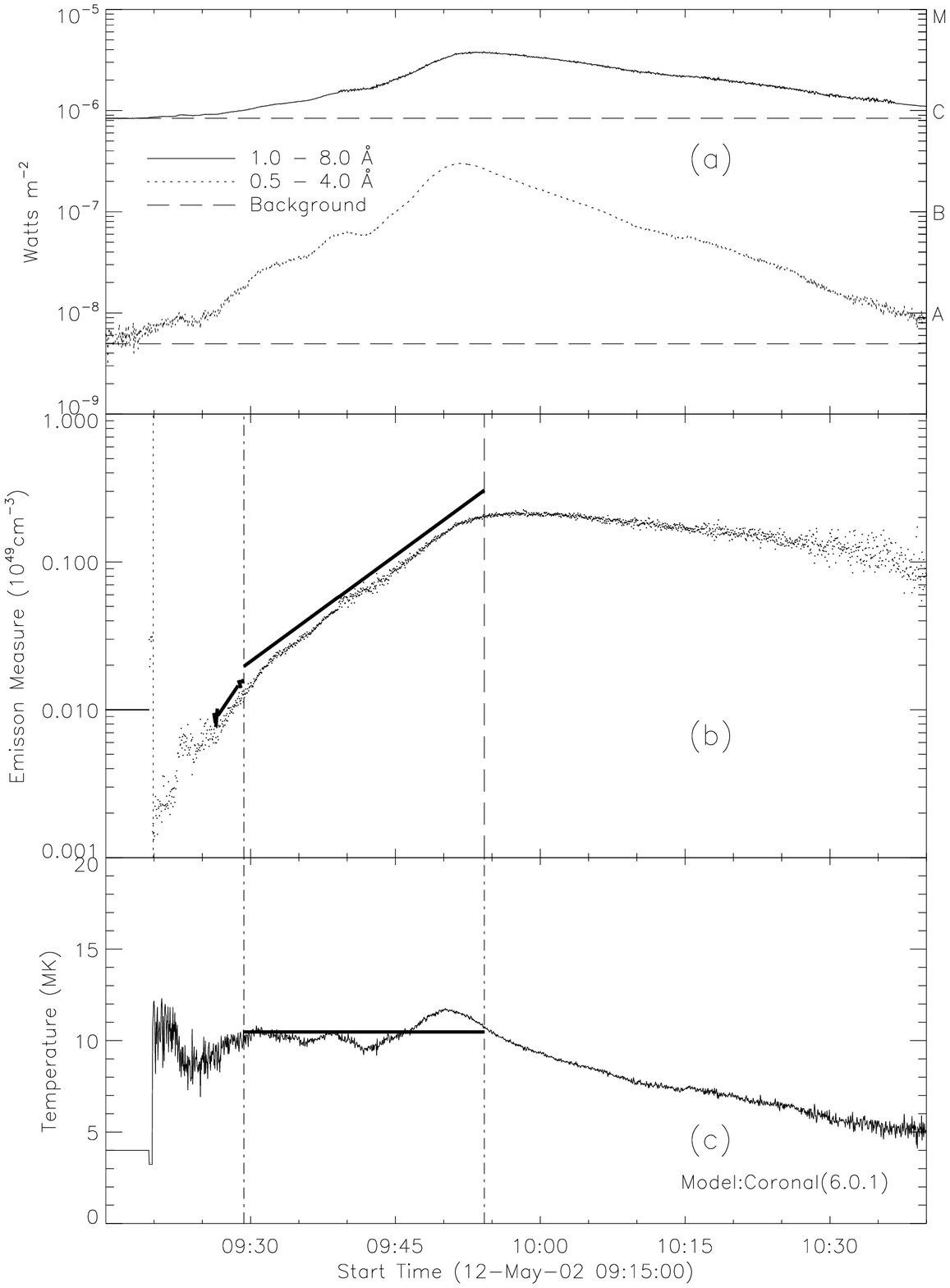}
\hfill
\includegraphics[width=8cm,height=13cm]{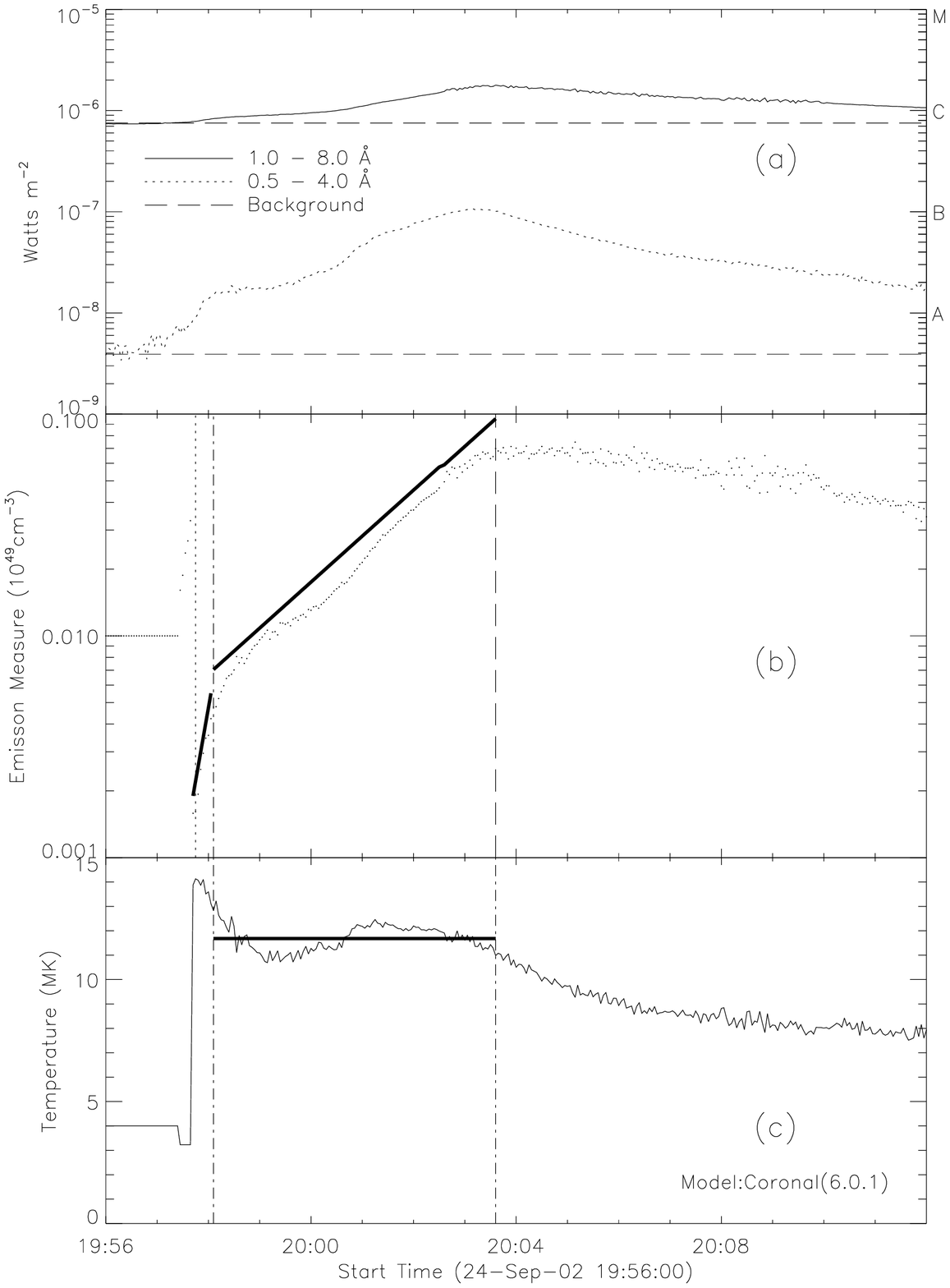}
\caption{Same as Figure \ref{lc} but for two flares with slightly more complicated temperature evolution when the EM grows exponentially. RHESSI does not have good coverage of these two flares.}
\label{lc1}
\end{figure}

Figure \ref{lc} shows two selected flares. The bottom panels show RHESSI light curves of these two flares. It is unfortunate that RHESSI did not cover the early rise phase of both flares. The flare in the left side is studied in detail by \citet{r09}, who showed that the flare is associated with a loop structure with prominent looptop and footpoint sources seen at different UV and EUV wavebands. The EM grows exponentially through the major part of the SXR rise phase. The temperature derived from GOES observation is nearly a constant in the rise phase.
There is no evidence of prominent impulsive hard X-ray emission near the flare peak time. The flare in the right side panel is very similar to the one in the left side except that there is evidence of impulsive emission above 25 keV near the flare peak time.

Figure \ref{lc1} shows two flares with slightly complicated light curves especially in the high energy channel during the dominant period of exponential growth in the EM. These complicated light curves lead to complicated behaviors in the inferred temperature evolution. These complexities may be attributed to fluctuations in the dominant process of exponential growth in the EM, and therefore these flares are considered to be similar to those in Figure \ref{lc}. However, the dominant period of exponential growth in the EM of these two flares extend to the flare peak time, which is different from the two flares in Figure \ref{lc}, where the dominant exponential growth period ends before the flare peak time.

From these analyses and for each selected flare, one can obtain the duration of the dominant exponential growth phase in the rise phase, the growth rate of the EM, the mean plasma temperature of this dominant exponential growth phase, the peak flux in the low energy channel, and the decay rate of the SXR flux in the low energy channel. In the following section, we will present the statistical properties of these quantities and their correlations.

\section{Results}
\label{res}

\begin{figure}[ht]
\includegraphics[width=8cm,height=4.944cm]{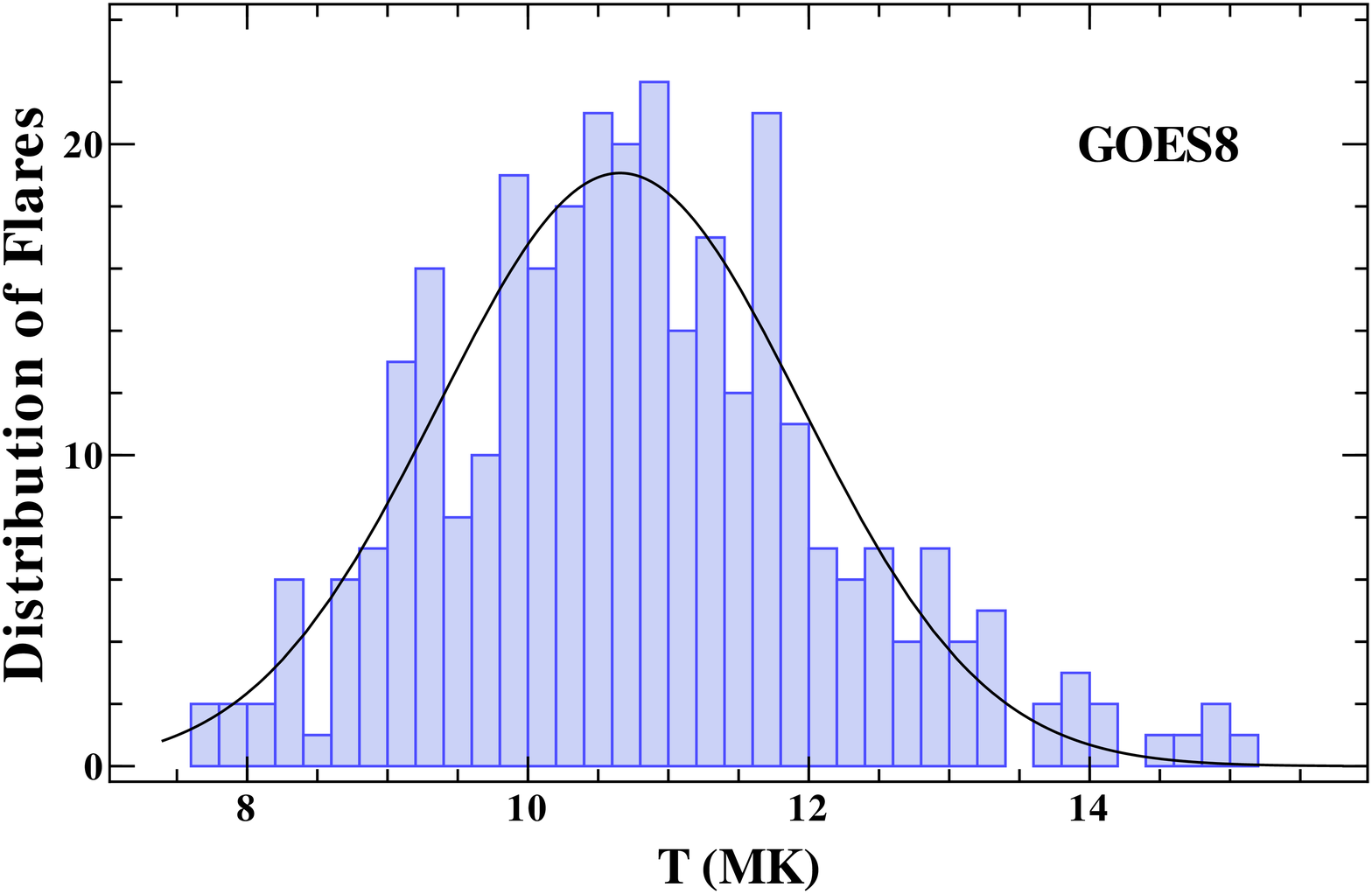}
\hfill
\includegraphics[width=8cm,height=4.944cm]{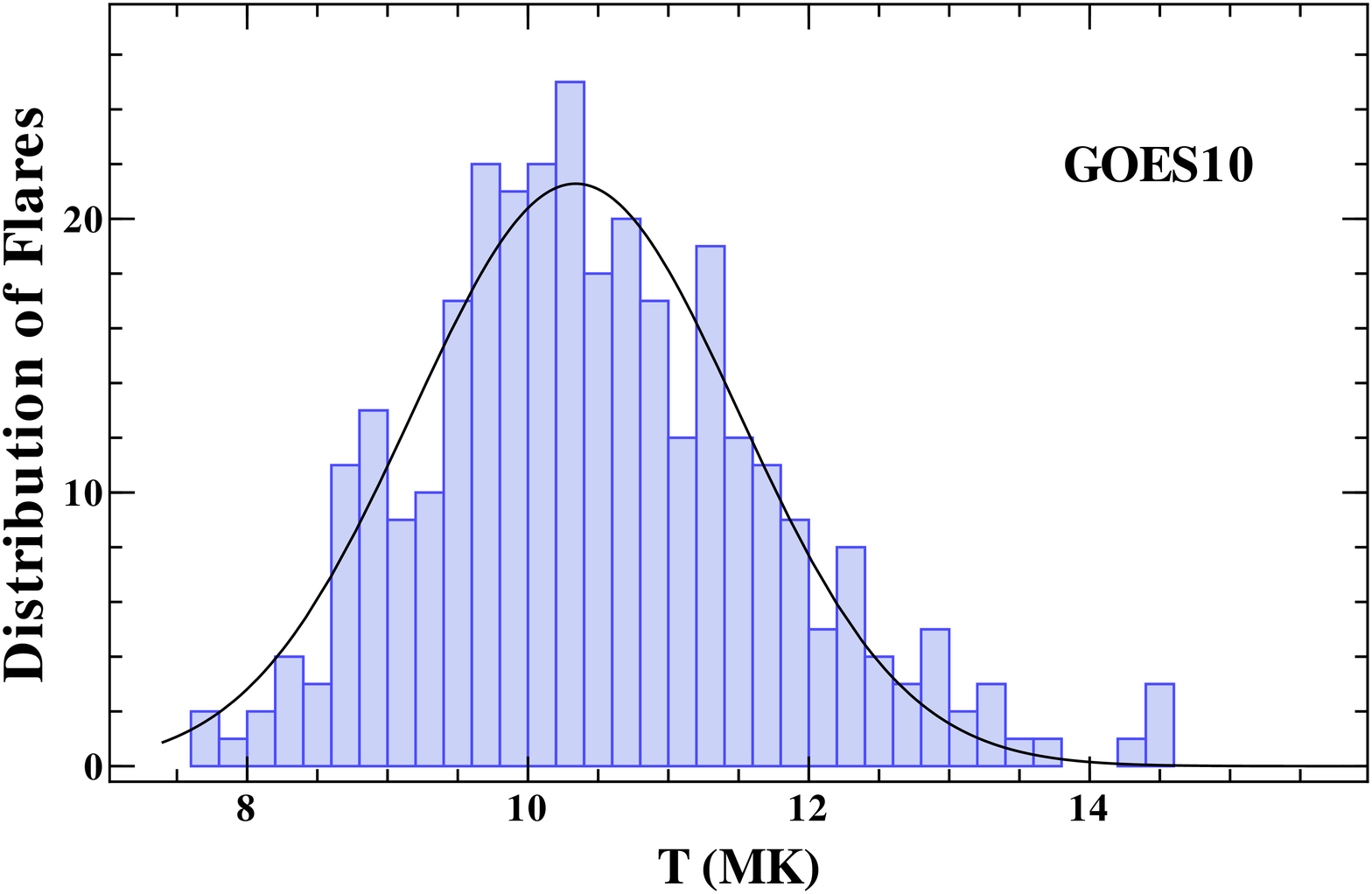}
\caption{The occurrence frequency distribution of the mean temperature of the dominant exponential growth phase of all flares selected from both satellite observations. The solid line shows a Gaussian fit with a mean of 10.7 MK and a standard deviation of 1.30 MK for GOES 8 (left) and 10.3 MK and 1.16 MK for GOES 10 (right).} \label{temper}
 \vspace{-0mm}
\end{figure}
With the above flare selection criteria, a total number of 620 and 522 flares are selected from GOES 8 and 10 observations, respectively. There are 316 flares selected from both satellite data. Figure \ref{temper} shows the occurrence frequency distribution of the mean temperature of the dominant exponential phase of the EM of these 316 flares. These distributions can be fitted with a normal distribution. The temperature measurement in the tails of this distribution may not be reliable. In the following, we will focus on flares within 1$\sigma$ range of these Gaussian distributions. There are 210 and 209 flares for GOES 8 and 10 observations, respectively, and 192 of them are identified from both satellite data. Table \ref{list} lists the characteristics of these 192 flares derived from GOES 8 observations.
Here the decay time is defined as the time it takes for the background subtracted SXR flux in the low energy channel to decrease by a factor of 2 from the peak value divided by $\ln({2})$. There are good agreements between GOES 8 and 10 observations. We therefore will not list the characteristics of these flares derived with GOES 10 observations.

\begin{center}
Table 1: Characteristics of 192 Selected Flares Derived from GOES 8
Observations
\include{longtable8}
\label{list}
\end{center}

\begin{figure}[ht]
\includegraphics[width=14cm]{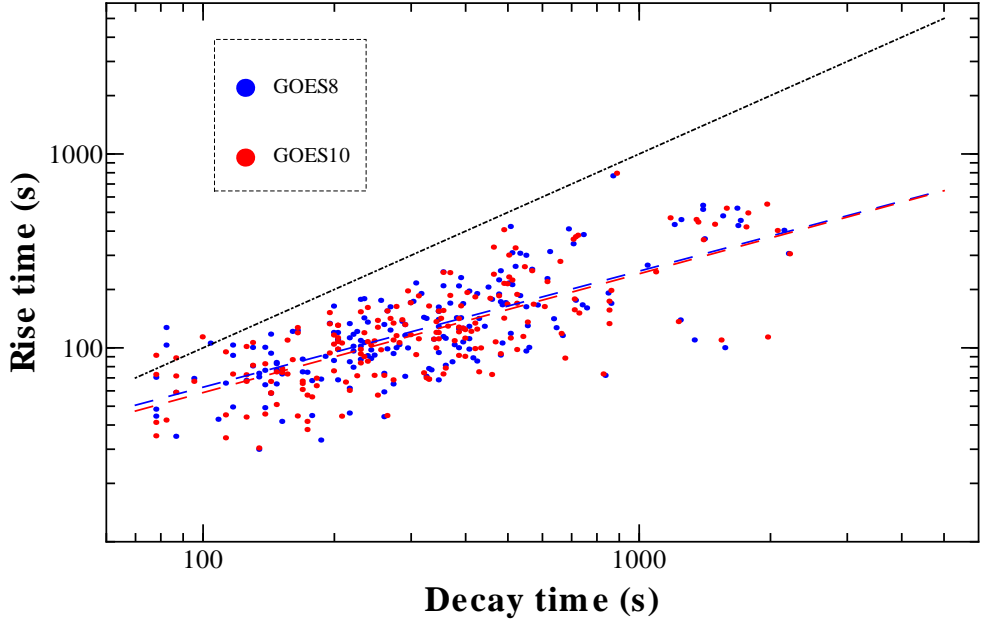}
\caption{
Correlation between the rise and decay times. The dashed lines indicate linear fits to the data derived from GOES 8 and 10 observations. The dot-dashed line indicates the equality of these two timescales.
} \label{decay}
 \vspace{-0mm}
\end{figure}
Figure \ref{decay} shows the correlation between the decay time of the SXR flux $t_d$ and the rise time of the dominant exponential growth period of the EM $t_e$ defined as the time required for the EM to grow by a factor of $e\simeq2.72$. For most flares, especially those with long decay time, the rise time is shorter than the decay time. Only for a few very short flares, the decay time is shorter than the rise time. The rise time increases slowly with the decay time. A linear fit to the correlation of the logarithm of these two timescales gives $t_e =4.0 (t_d/s)^{0.60}s$ and $t_e=3.5 (t_d/s)^{0.61}s$ for GOES 8 and 10 observations, respectively.

\begin{figure}[ht]
\subfigure[]{
\begin{minipage}[b]{0.5\textwidth}
\includegraphics[width=8cm,height=4.944cm]{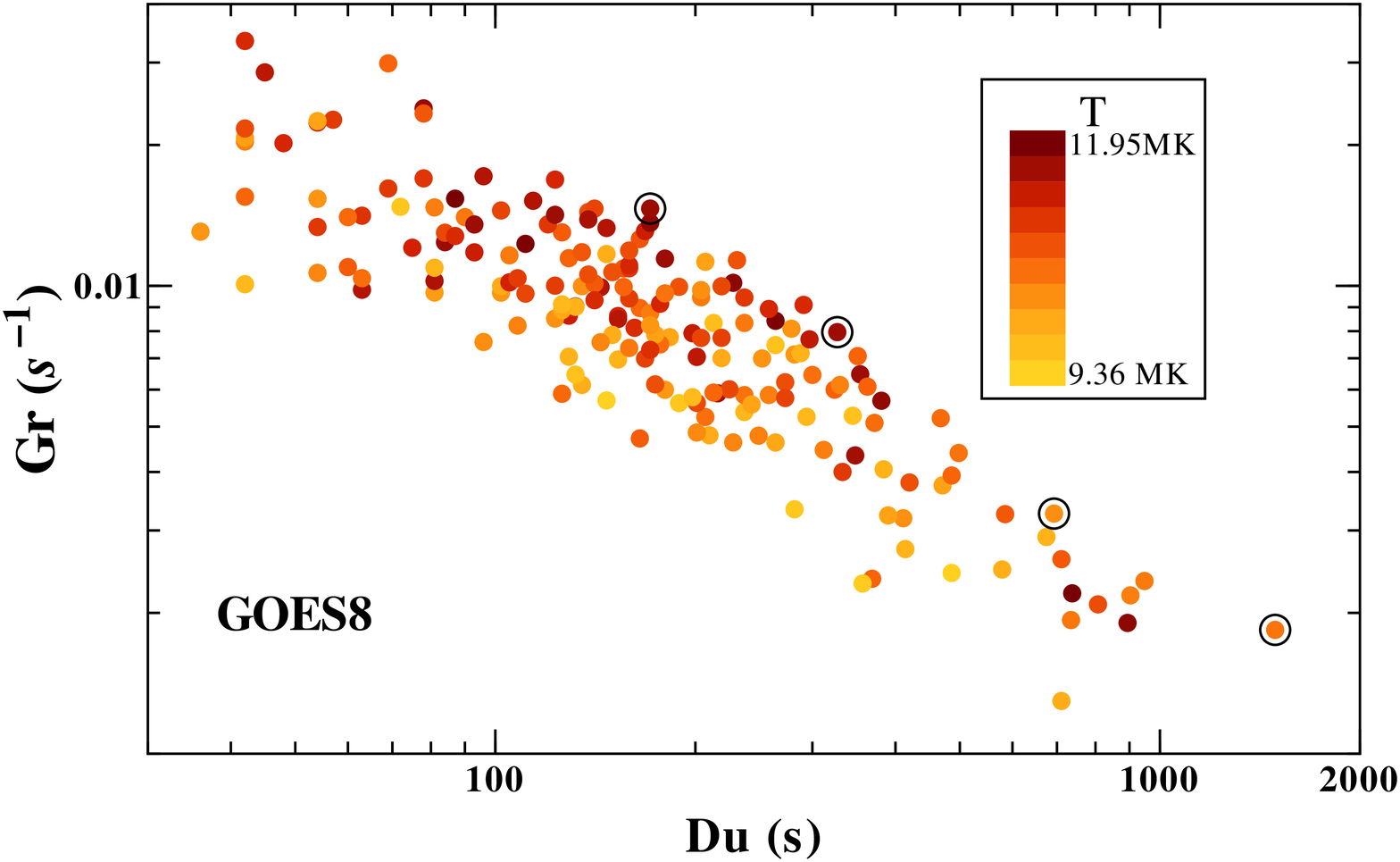}
\includegraphics[width=8cm,height=4.944cm]{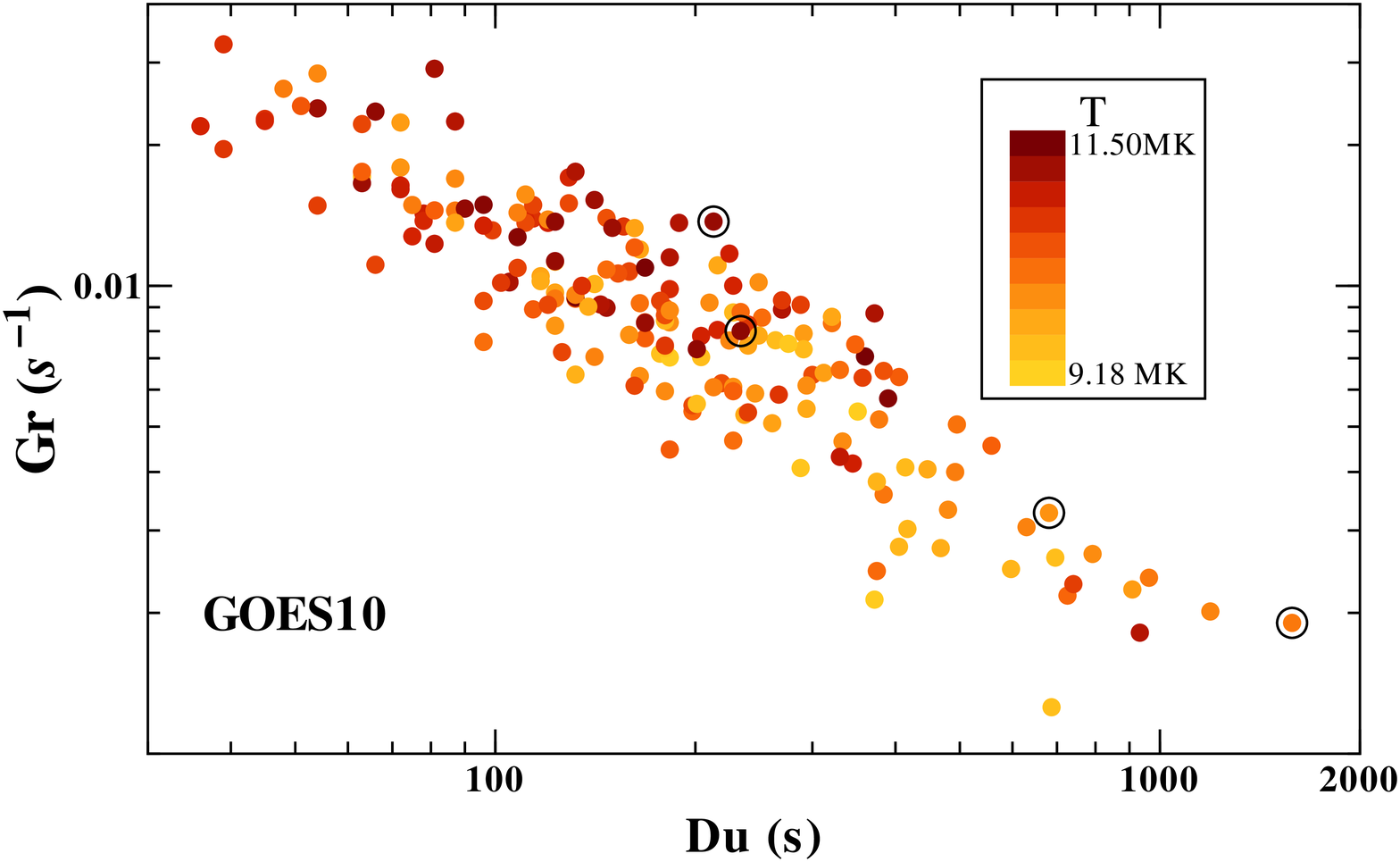}
\end{minipage}
}
\hfill
\subfigure[]{
\begin{minipage}[b]{0.5\textwidth}
\includegraphics[width=8cm,height=4.944cm]{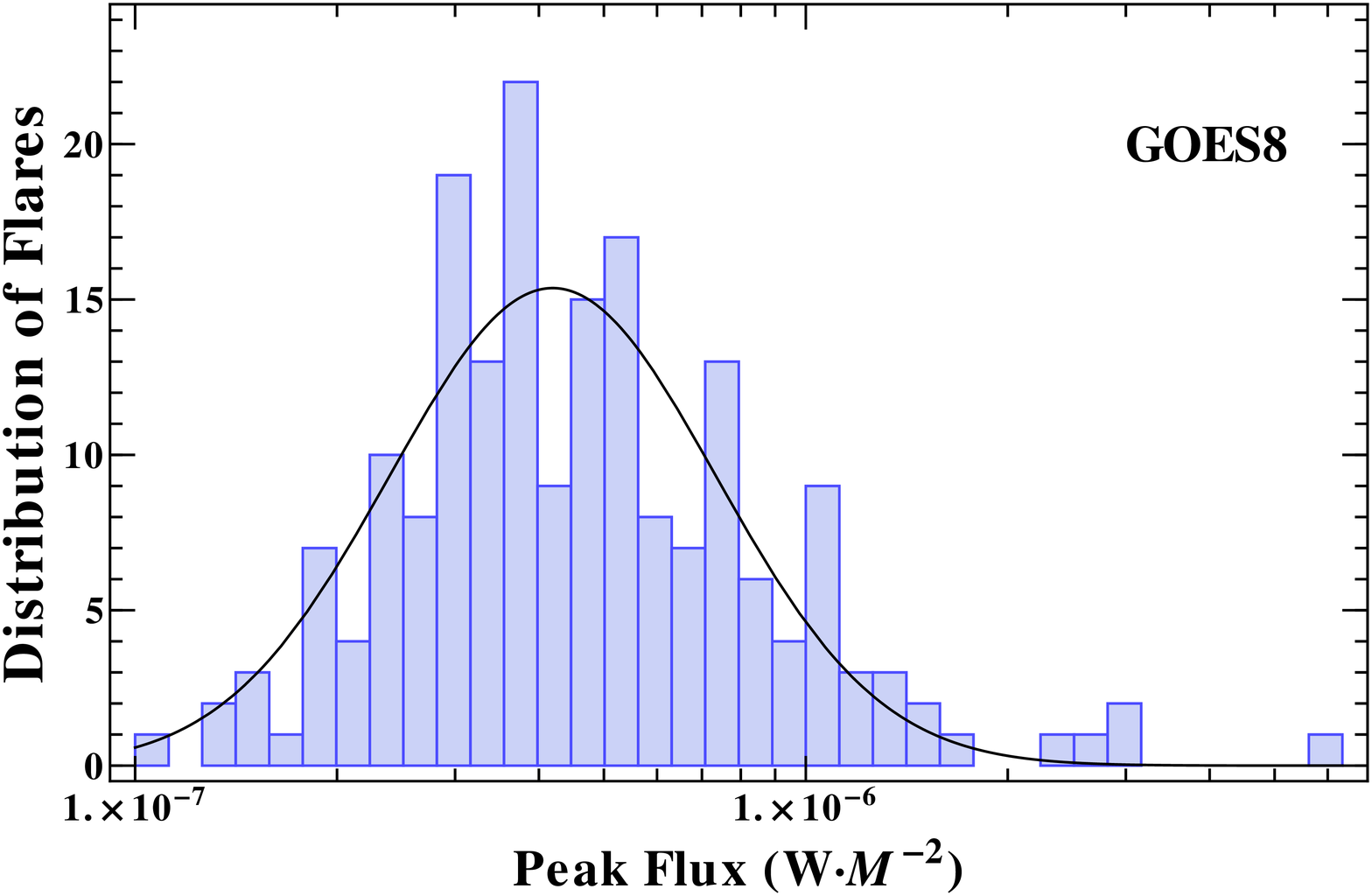}
\includegraphics[width=8cm,height=4.944cm]{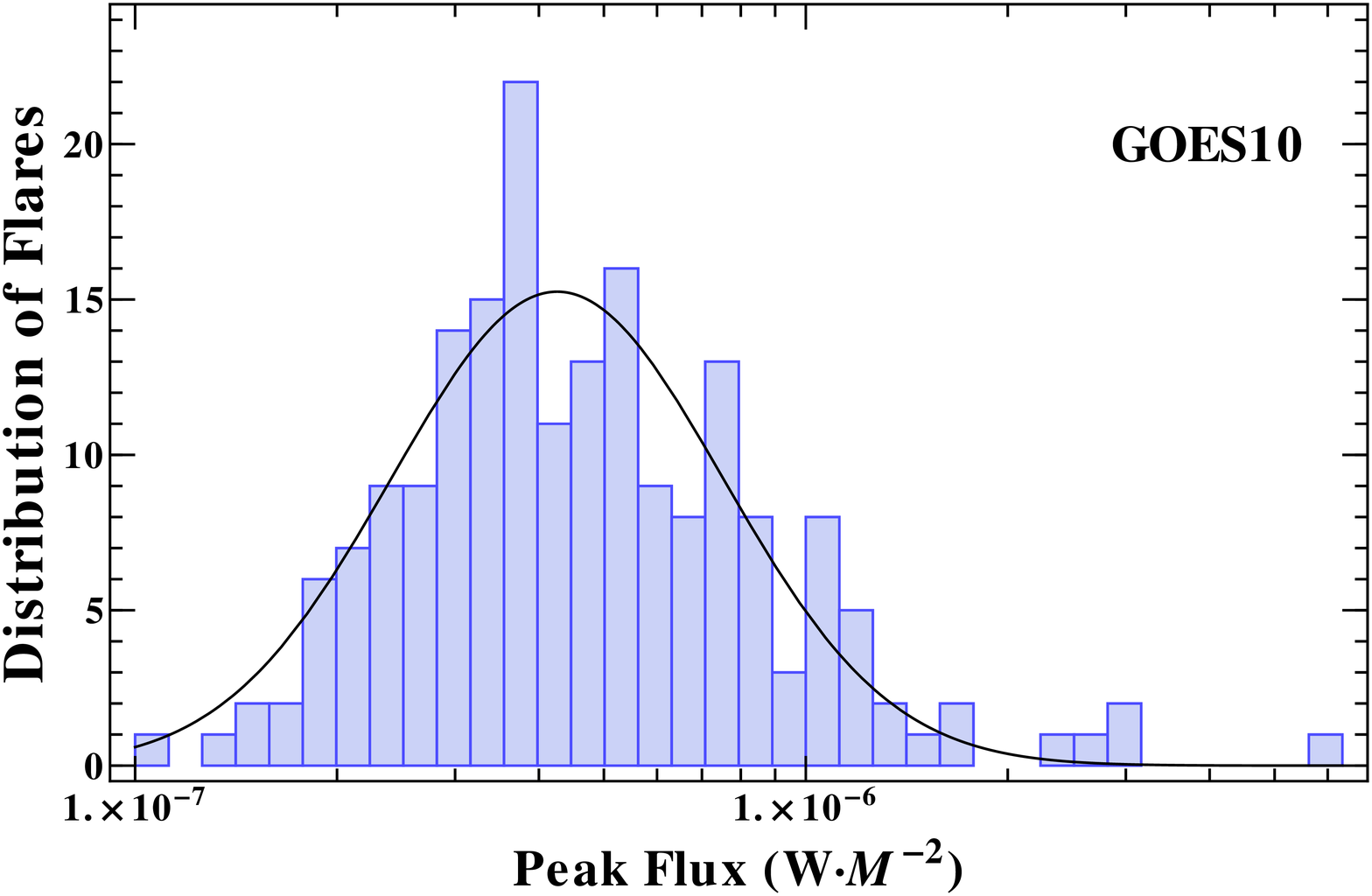}
\end{minipage}
}
\caption{Left: Correlation between the duration of the dominant exponential growth phase $Du$ and the corresponding growth rate $Gr$. The mean
temperature is indicated by the color bar. The circles indicate the four flares shown in Figures \ref{lc} and \ref{lc1}.
 Right: Distribution of the background-subtracted peak flux of the 1 - 8 \AA\  band of the selected flares. The solid line shows a log-normal fit with a mean of  $10^{-6.4}$ W M$^{-2}$ and a standard deviation of ${0.24}$ for both GOES 8 and 10 observations.
} \label{result}
 \vspace{-0mm}
\end{figure}
The most unexpected finding of this study is a strong anti-correlation between the growth rate $Gr=t_e^{-1}$ and the duration of the dominant exponential growth period $Du$ as shown in the left panels of Figure \ref{result}. The result also indicates that $Gr$ increases with the increase of the mean temperature $T$. A linear fit to the correlation among $\log(Gr)$, $\log(Du)$ and $\log(T)$ gives
$$\log(Gr/{\rm s}^{-1}) = -0.69\log(Du/{\rm s}) + 1.9 \log(T/{\rm MK}) - 2.5$$
and
$$\log(Gr/{\rm s}^{-1}) = -0.73\log(Du/{\rm s}) + 1.5 \log(T/{\rm MK}) - 2.0$$
for GOES 8 and 10 observations, respectively. Such an anti-correlation suggests that the EM stop to increase exponentially after reaching certain level. Indeed, the occurrence frequency distribution of the peak flux of the selected flares shows a relatively narrow log-normal distribution\footnote{Given the relatively small sample size of the selected flare, we can not tell whether this distribution is consistent with the peak flux distribution of all flares shown in the left panels of Figure \ref{bk}.} as shown in the right panels of Figure \ref{result}. Since the temperature covers a narrow range,
the SXR peak flux gives a rough measurement of the EM at the peak time. The observed anti-correlation between $Gr$ and $Du$ therefore is consistent with the relatively narrow distribution of the SXR peak flux.
All of these selected flares belong to GOES class B or C. Big flares are likely more complex and therefore have less chance to meet our selection criteria. However, there is a slight excess relative to the log-normal distribution at high values of the peak flux.

\begin{figure}[ht]
\includegraphics[width=8cm,height=4.944cm]{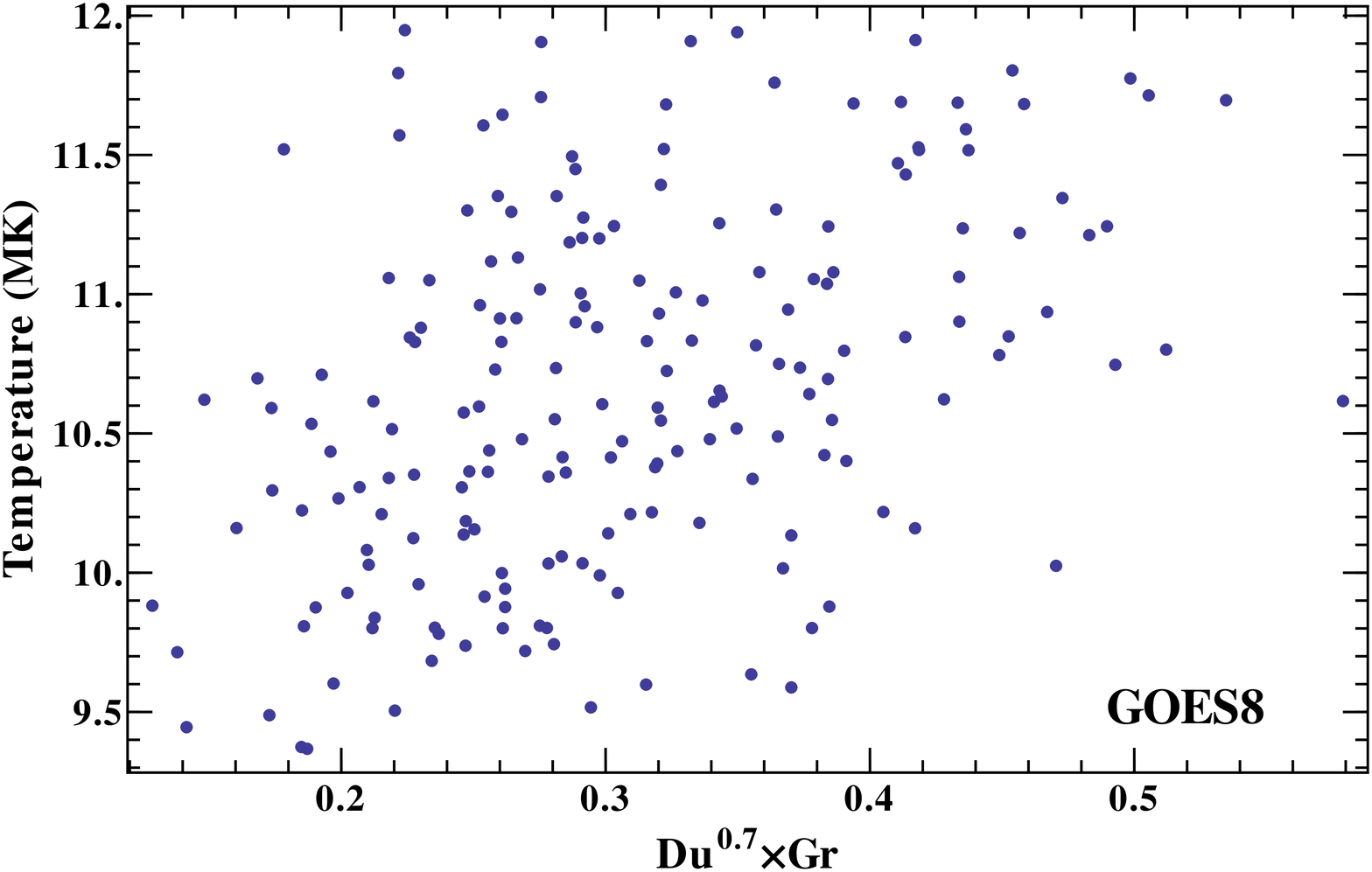}
\hfill
\includegraphics[width=8cm,height=4.944cm]{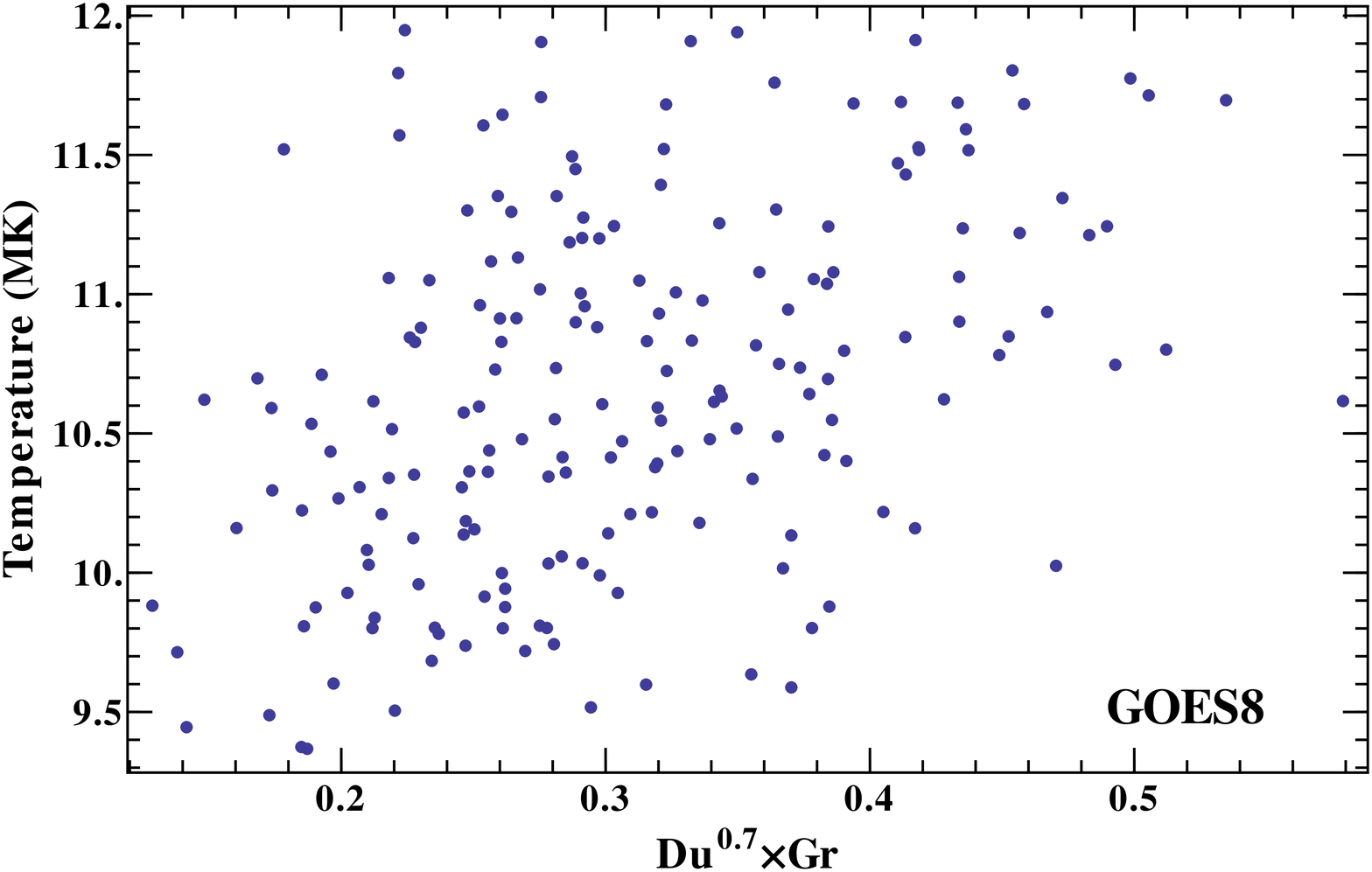}
\caption{Correlation between the mean temperature $T$ and $Du^{0.7}Gr$.} \label{temp}
 \vspace{-0mm}
\end{figure}
The correlation between $Gr$, $Du$ and $T$ is much weaker. Based on the fitting result of their correlation above, Figure \ref{temp} shows the correlation between $Du^{0.7} Gr$ and $T$. Although $Du^{0.7}Gr$ tends to increase with the increase of $T$, the spread of the correlation is big. Therefore quantitative dependence of the results on $T$ may not be trust worthy.

\begin{figure}[ht]
\subfigure[]{
\begin{minipage}[b]{0.5\textwidth}
\includegraphics[width=8cm,height=4.944cm]{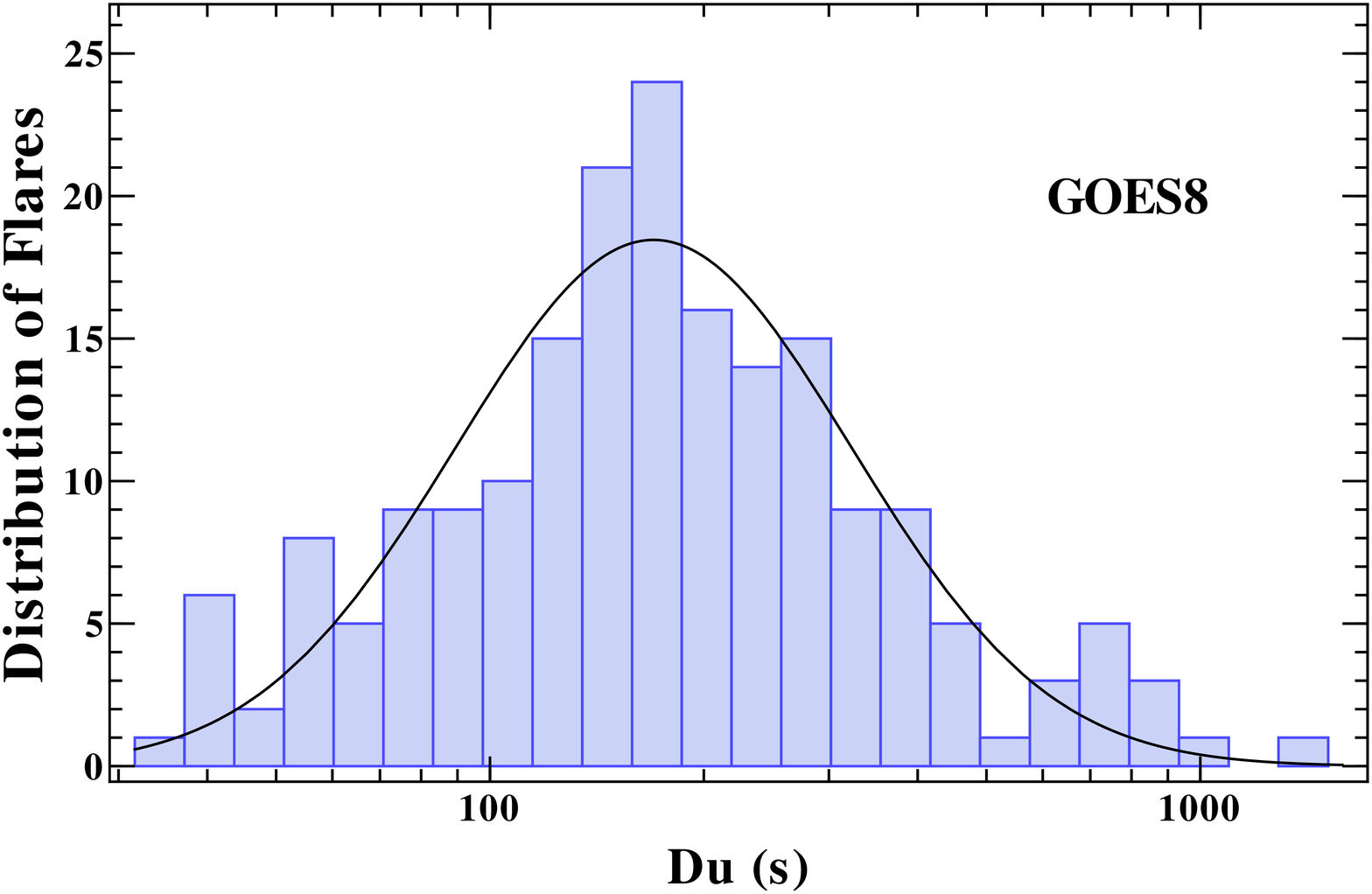}
\includegraphics[width=8cm,height=4.944cm]{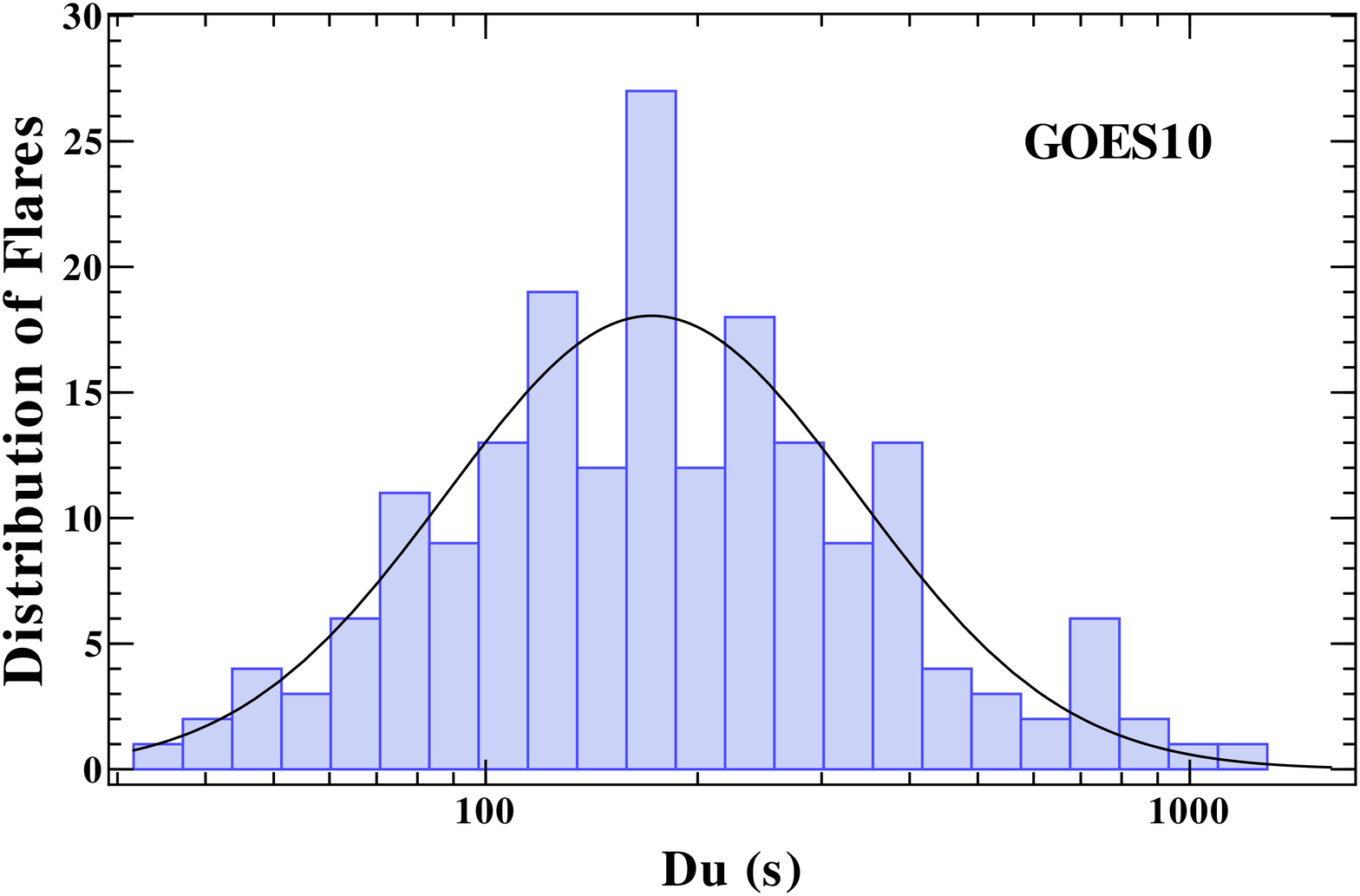}
\end{minipage}
}
\hfill
\subfigure[]{
\begin{minipage}[b]{0.5\textwidth}
\includegraphics[width=8cm,height=4.944cm]{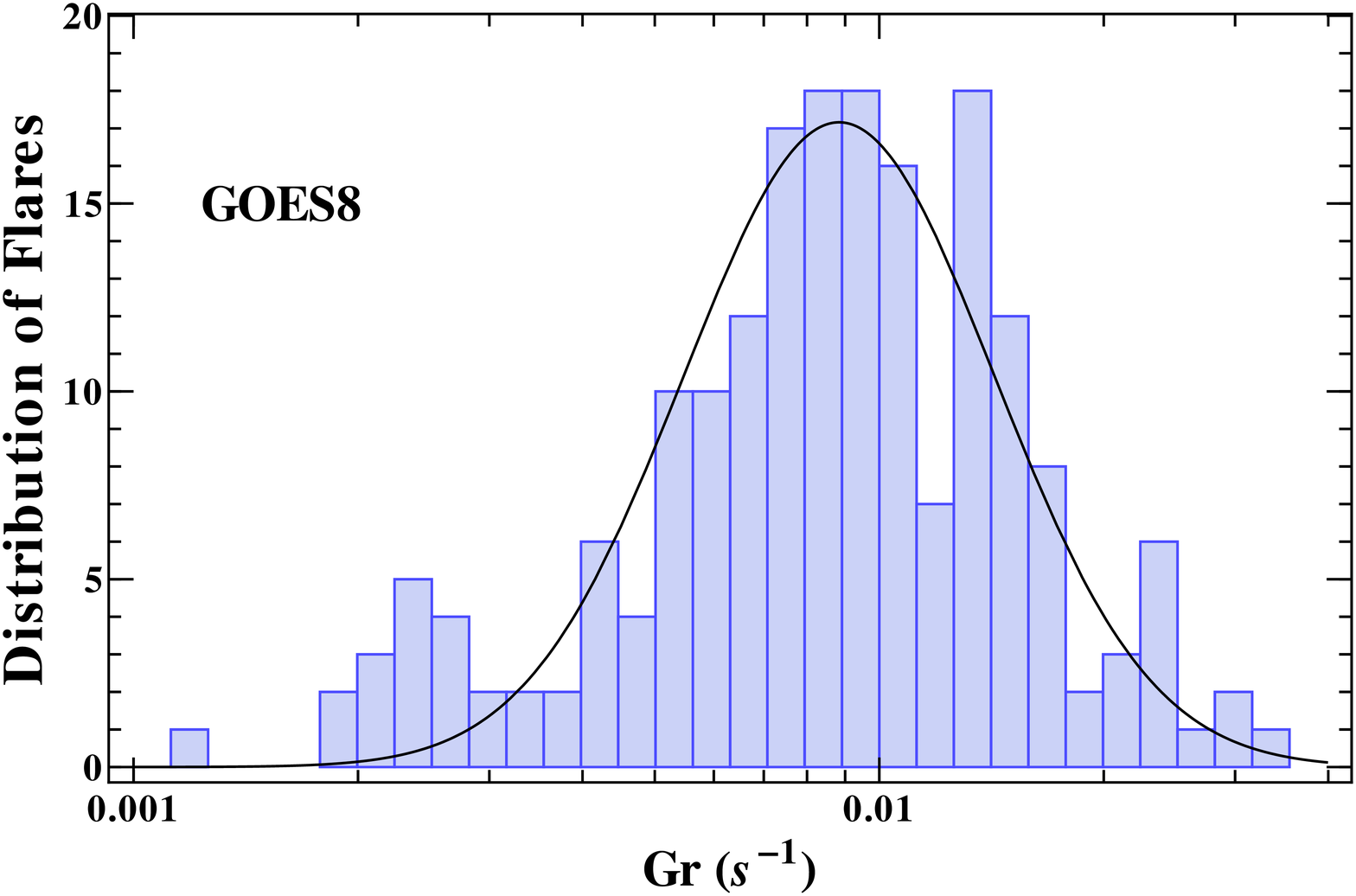}
\includegraphics[width=8cm,height=4.944cm]{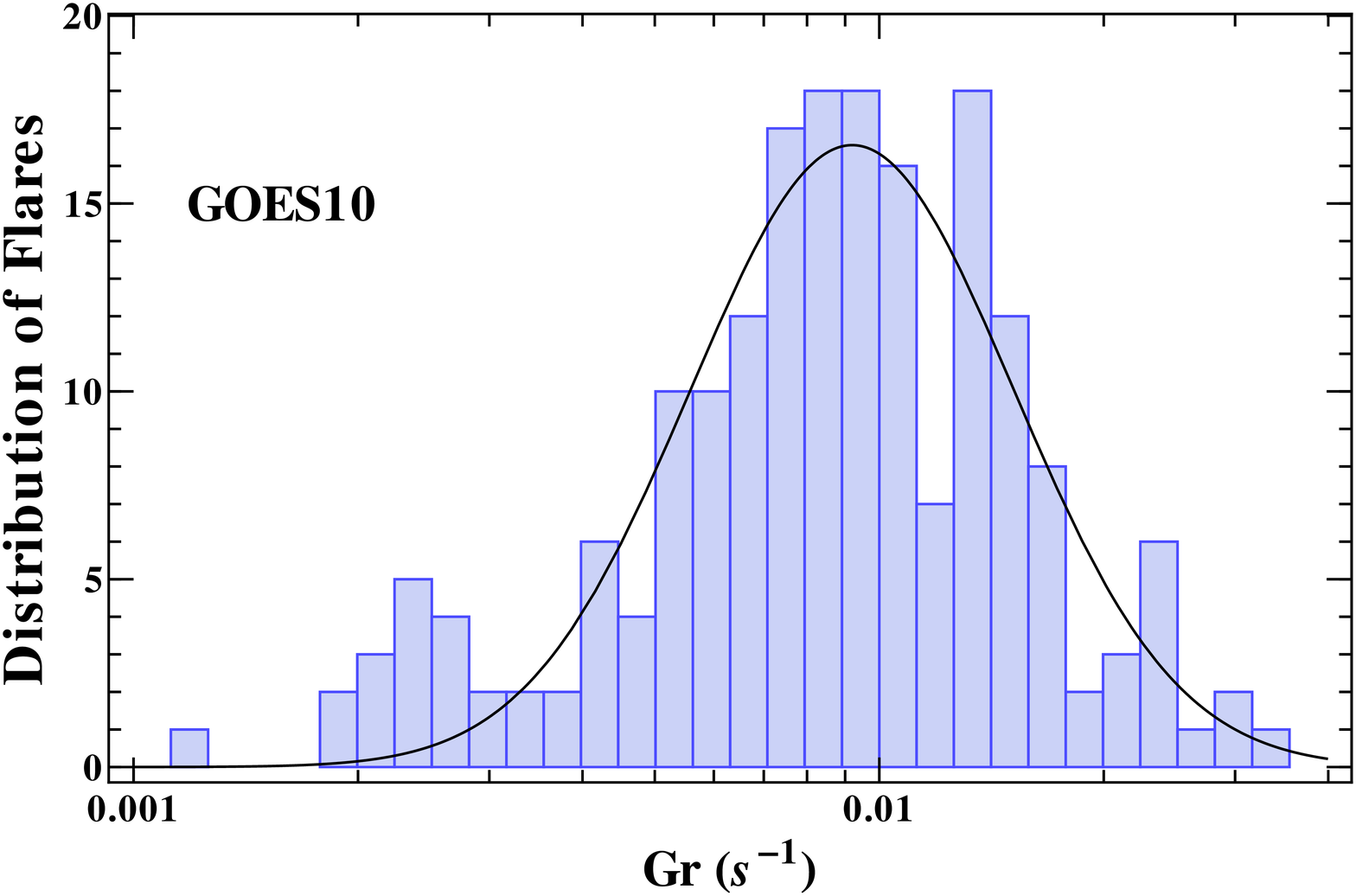}
\end{minipage}
}
\caption{Left: Same as the right panels of Figure \ref{result} but for the
 frequency distribution of the duration of the dominant exponential growth phase. The log-normal fit has a mean of  $170$s and a standard deviation of $1.9$ for both GOES 8 and 10.
 Right: Same as the left panels but for the distribution of the growth rate of the exponential growth phase.  The log-normal fit has a mean of  $0.009$ s$^{-1}$ and a standard deviation of $1.6$ for both GOES 8 and 10.
} \label{durgrow}
\end{figure}
Figure \ref{durgrow} shows the occurrence frequency distribution of the growth rate and duration of the dominant exponential growth period. Both distributions can be fitted with a log-normal function. One of our selection rules requires the duration being longer than 30 s, which explains the low bound of this quantity. The longest duration of the exponential growth period is about half an hour. Relative to the log-normal distribution, the obtained occurrence frequency distributions also have slight excesses at longer durations and lower growth rates. Selection of more similar flares from other observation period may shade light on the significance of these excesses.

\section{Discussion and Conclusions}
\label{con}

To uncover the dominant physical processes in flaring loops and give more quantitative modeling, we have obtained a sample of flares with relatively simple SXR light curves from GOES observations. The complexity of the flare phenomena caused by the complex coronal environment is partially suppressed via our selection of flares with relatively simple time evolution. Specifically, we have focused on a class of flares whose SXR rise phase is dominated by a period of exponential growth of the EM. Detailed multi-wavelength studies show some of these flares are associated with single loops.

The rise time ranges from 30 s to more than 10 minutes suggesting a (magneto-) hydrodynamical process. There are two possible mechanisms that can lead to a period of exponential growth of the EM. If the loop structure is relatively simple and stable, the increase of the EM has to be caused by the evaporation of plasmas from the chromosphere. The exponential growth of the EM implies exponential growth of the thermal energy and therefore a heating rate proportional to the thermal energy density. The latter suggests a feedback of the heated plasma on the energy dissipation processes. Since it is commonly accepted that the flare energy release happens in the corona, the evaporation has to be driven by energy fluxes from the corona to the chromosphere. The fact that the heating rate is proportional to the thermal energy density implies a saturated energy flux from the corona to the footpoints. Such a saturated energy flux may be caused by the saturated conduction in a low density plasma caused by the non-local transport of energetic particles in the loop \citep{j06, b09}. In such a scenario, we would expect strong emission from the footpoints.

If the topological structure of the magnetic fields in the loop is complex, for example, the magnetic field lines may be twisted and braided \citep{w10}, the flare may be associated with a filament structure \citep{l09}. The exponential growth of the EM can be caused by an exponential growth of the volume filling factor of the heated plasma in the filament. Strong evaporation from the chromosphere is not necessary in such a case. More detailed multi-wavelength studies of individual flare are necessary to distinguish the two scenarios. From the smooth distribution of the characteristics of the selected flares, the two scenarios are not distinguishable from the SXR light curves alone.

We find a strong anti-correlation between the growth rate of the EM and the duration of the dominant exponential growth period, which suggests that the exponential growth phase ends when the EM reaches certain level. The ending of the exponential growth phase implies a new phase of energy release in the impulsive phase. According to the two mechanisms proposed above, the ending can be caused by either a high density of evaporated plasmas or the volume filling factor reaching a saturation level. Although there is no evident reason why the EM stops to grow exponentially once reaching certain level, the observed anti-correlation is consistent with the relative narrow log-normal distribution of the peak flux of the selected flares. Most of the selected flares belong to GOES B with a small fraction belonging to GOES C. The relatively small amplitude of these elected flares can be partially attributed to more complexity of bigger flares.

The peak flux of flares in general follows a power-law distribution \citep{v02a}. The distribution of the peak flux of the selected flares, the duration of the exponential growth period, and the growth rate, however, follow a relatively narrow log-normal distribution. While a power-law distribution implies a lack of characteristic scales in the system. Our relatively narrow distribution of selected flares suggests that they may represent a particular class of flares with a characteristic peak flux of $\sim4\times 10^{-7}$ W M$^{-2}$. The selected flares have relative simple light curves and are likely associated with single loops. If these results are further confirmed with a larger flare sample, we can describe flares in general as a set of flaring loops. The power-law distribution of the flare peak flux is mostly caused by dramatic variation of the number of loops in different flares. For an X class flare, thousands of loops should be activated. Therefore the flare study may be separated into two aspects: 1) physical processes in a flaring loop; 2) the topological structure of the flare region that determines the number of loops to be activated during a flare.

The flares selected here show relatively gradual evolution in general and the impulsive hard X-ray (HXR) emission is relatively weak, similar to the slow long-duration events studied by \citet{b11}. The more gradual evolution of the SXR may be intimately connected to the lack of impulsive HXR emission. Earlier studies by \citet{s06} suggest that the impulsive HXR emission is better correlated with the SXR growth rate than with the SXR flux. Since the emission is dominated by the gradual emission component in the rise phase, the process of particle acceleration may be unimportant. Flares with prominent particle acceleration may correspond to a class of events distinct from
the flares selected here. Most observed characteristics of the selected flares should be explained in the context of magneto-hydrodynamic evolution of flaring loops \citep{w10}.

\acknowledgements
 We thank Youping Li, Hugh Hudson, Lyndsay Fletcher, and Peter J. Cargill for helpful discussions. This work is supported by the National Natural Science Foundation of China via grants 11143007 \& 11173064.

{}

\end{document}

%% file: longtable8.tex
\begin{longtable}{c r r r r r}
\label{longtab:list} \\
\hline
Peak Time & Peak Flux & Duration & Decay Time & Growth Rate & Temperature \\
\hline
UT & $10^{-7}$W m$^{-2}$ & s & s & $10^{-3}$ s$^{-1}$ & MK \\
\hline
\endfirsthead
\hline
Peak Time & Peak Flux & Duration & Decay Time & Growth Rate & Temperature \\
\hline
\endhead
\endfoot
\hline
\endlastfoot
1999-01-12T20:44:27&7.93& 153& 201&3.74&11.28\\
1999-01-27T01:21:39&12.15& 282& 240&3.10&10.13\\
1999-01-31T11:43:39&6.17&  84&  90&5.38&11.71\\
1999-01-31T16:28:03&4.81&  42&  96&8.83&10.35\\
1999-02-01T23:31:33&8.79&  87& 189&6.67&11.94\\
1999-02-02T05:10:03&6.03&  81&  57&4.20&10.08\\
1999-02-11T23:15:27&13.49& 579&1488&1.08&9.84\\
1999-02-18T17:20:30&21.01& 324& 513&2.61&10.65\\
1999-02-22T01:38:30&7.49& 126& 255&2.55&10.59\\
1999-02-25T13:42:15&4.61&  63&  93&6.13&11.12\\
1999-03-03T07:57:03&7.34& 129& 159&4.98&10.63\\
1999-03-03T13:25:12&8.38& 159& 132&4.80&11.24\\
1999-03-03T19:47:27&7.20&  78& 105&10.40&11.71\\
1999-03-09T23:27:09&8.61& 129& 462&3.75&11.35\\
1999-03-12T13:29:27&8.21&  48&  81&8.76&11.24\\
1999-03-12T16:37:00&13.53& 138&  66&6.24&10.85\\
1999-03-15T02:01:21&8.47& 153&  99&3.69&11.49\\
1999-03-15T15:52:51&7.37& 252& 270&3.04&10.18\\
1999-03-17T03:42:42&11.57& 147& 117&5.77&11.52\\
1999-04-06T02:59:33&6.00& 144&  96&4.31&11.52\\
1999-04-06T04:17:09&7.92&  81& 168&4.45&11.57\\
1999-04-06T07:05:42&36.13& 948&1167&1.02&10.41\\
1999-04-07T05:36:42&12.47& 159& 162&4.08&11.01\\
1999-04-09T04:15:39&6.20& 123& 153&3.70&10.19\\
1999-04-09T15:54:27&12.94& 372& 282&2.21&10.55\\
1999-04-09T20:09:39&5.43&  42&  54&8.99&10.06\\
1999-04-11T02:59:09&4.21&  42&  96&6.73&10.62\\
1999-04-13T09:21:21&6.37& 126& 102&5.65&10.70\\
1999-04-18T21:19:24&7.88& 264& 213&2.01&9.96\\
1999-04-21T00:22:36&6.05& 135& 258&5.12&10.75\\
1999-04-21T11:56:30&6.00& 120& 180&5.88&11.08\\
1999-04-21T14:06:54&7.61& 111& 249&5.34&11.91\\
1999-04-24T18:23:24&3.47& 228& 348&2.01&10.31\\
1999-04-27T09:37:15&5.71& 204& 306&3.36&10.93\\
1999-05-03T18:14:51&7.61&  75& 102&5.24&11.30\\
1999-05-04T11:29:09&9.48& 135& 267&4.34&10.21\\
1999-05-25T18:10:12&16.17& 585& 369&1.41&10.73\\
1999-05-26T14:12:03&10.42&  93&  93&5.87&11.68\\
1999-05-29T09:34:33&11.46& 201& 159&2.44&10.88\\
1999-05-31T15:46:54&6.60&  60& 165&4.76&10.71\\
1999-05-31T20:26:36&14.03& 354& 240&2.81&11.68\\
1999-06-09T00:18:12&12.70& 471& 723&1.63&10.03\\
1999-06-09T17:12:36&13.00& 216& 597&2.56&11.61\\
1999-06-11T21:36:42&13.96& 165& 153&5.47&10.78\\
1999-06-12T23:51:39&21.06& 228& 282&4.41&11.80\\
1999-06-20T00:59:42&20.14& 156& 333&4.73&10.74\\
1999-06-23T15:59:09&14.11& 213& 138&3.62&9.63\\
1999-07-07T14:50:27&12.59& 183& 240&3.37&9.99\\
1999-07-10T05:10:15&10.50& 168& 105&5.69&11.35\\
1999-07-10T15:12:39&8.80&  90& 225&6.09&10.44\\
1999-07-12T16:47:57&13.28& 207& 288&4.89&10.02\\
1999-07-15T15:58:42&10.29& 258& 342&2.54&10.36\\
1999-07-16T00:49:27&5.08&  54&  99&4.63&10.30\\
1999-07-16T07:40:36&7.93& 102&  81&4.20&10.14\\
1999-07-21T14:06:15&7.74& 102& 192&4.33&9.91\\
1999-08-13T11:29:54&6.77& 207& 219&2.28&10.52\\
1999-08-13T23:34:39&6.01&  36&  96&5.67&10.16\\
1999-09-10T07:43:36&7.13& 177& 150&3.26&10.55\\
1999-09-10T09:20:24&8.82& 300& 183&2.80&10.52\\
1999-09-11T14:44:15&8.15& 165& 282&3.89&10.59\\
1999-09-20T00:58:27&13.33& 486& 393&1.71&10.61\\
1999-09-22T09:10:57&9.77&  96& 201&3.29&10.22\\
1999-09-29T15:23:45&7.00& 411& 432&1.38&10.21\\
1999-09-30T23:46:24&7.51& 219& 282&3.36&10.98\\
1999-10-01T03:19:27&10.21& 171& 246&3.80&10.39\\
1999-10-03T14:28:51&9.11& 180& 336&2.60&10.12\\
1999-10-13T04:53:18&11.76& 105& 141&4.42&11.30\\
1999-10-15T07:20:36&13.10&  69& 129&12.98&10.62\\
1999-10-18T02:31:24&18.32& 291& 927&3.95&11.21\\
1999-11-01T22:35:30&9.60& 159& 171&4.73&11.05\\
1999-11-02T09:49:03&14.57& 384& 246&1.76&9.80\\
1999-11-30T11:12:39&12.85& 198& 273&3.44&11.39\\
1999-12-05T03:27:54&9.30& 294& 294&2.28&9.74\\
1999-12-06T09:51:15&12.23& 390& 354&1.40&10.03\\
1999-12-14T07:00:03&11.73& 237& 369&2.33&9.74\\
2000-01-05T04:15:30&10.62& 369& 351&1.03&10.62\\
2000-01-07T15:16:03&11.59& 237& 207&2.54&10.48\\
2000-01-10T19:02:45&13.04& 675& 489&1.26&9.80\\
2000-01-25T11:37:00&6.51& 147& 177&2.47&9.37\\
2000-01-29T00:30:51&7.90& 189& 318&2.44&9.50\\
2000-01-29T12:53:57&5.92& 108& 111&3.57&10.34\\
2000-02-05T18:20:33&12.73& 288& 861&3.12&9.80\\
2000-02-29T21:58:54&18.72& 150& 189&4.65&10.82\\
2000-04-16T18:02:48&8.55&  81&  81&4.75&9.78\\
2000-04-26T12:33:15&11.02& 123& 180&7.32&11.24\\
2000-04-27T03:20:27&9.39&  81&  87&6.39&10.38\\
2000-04-27T15:50:36&8.99& 147& 294&5.08&9.88\\
2000-04-30T16:18:48&8.33& 174& 240&2.67&10.83\\
2000-04-30T16:38:18&7.68&  93& 150&5.12&11.35\\
2000-05-03T21:47:09&11.45& 237& 141&3.62&10.42\\
2000-05-09T04:19:54&11.17& 357& 834&1.00&9.45\\
2000-05-28T07:25:12&16.56& 279& 186&3.52&10.16\\
2000-05-29T12:29:36&10.10& 231& 168&4.93&10.80\\
2000-05-30T18:47:06&8.13& 123& 204&4.35&11.00\\
2000-06-19T05:31:36&11.62& 132& 141&3.92&11.91\\
2000-06-21T00:46:54&67.11& 264& 351&3.66&11.91\\
2000-06-26T04:35:15&19.27& 381& 495&2.47&11.76\\
2000-07-03T04:15:12&8.19& 282& 381&1.45&9.49\\
2000-07-03T16:29:36&7.58&  42& 150&9.42&10.88\\
2000-08-18T23:55:21&13.39& 363& 138&2.64&10.64\\
2000-08-23T01:16:00&14.08& 711& 603&0.56&9.88\\
2000-08-24T06:42:12&18.14& 903& 864&0.95&10.36\\
2000-08-28T11:42:42&11.71& 204&  72&4.11&10.40\\
2000-09-04T15:37:42&14.03& 171& 105&5.92&11.77\\
2000-09-14T13:56:18&22.60& 258& 249&3.88&11.24\\
2000-10-17T11:22:21&11.13& 312& 330&1.94&10.36\\
2000-10-22T17:04:24&8.50& 108& 381&4.51&11.02\\
2000-11-10T16:39:15&18.07& 330& 381&2.67&10.34\\
2000-11-16T06:55:27&9.61&  54& 123&9.70&11.30\\
2000-11-17T02:23:30&13.94& 159& 102&5.17&10.85\\
2000-11-21T16:48:18&11.94& 156& 387&4.32&10.61\\
2000-12-07T02:45:12&10.56&  96&  99&7.45&11.52\\
2000-12-13T17:01:39&11.62& 150&  57&3.41&9.88\\
2000-12-14T18:28:57&15.03& 204& 159&4.25&10.22\\
2000-12-25T08:46:39&9.66& 105& 138&5.05&10.41\\
2000-12-26T04:16:39&11.64&  84& 231&5.64&10.90\\
2001-01-11T03:48:36&13.56& 219& 162&3.04&9.93\\
2001-01-15T20:24:00&13.24& 273& 210&2.50&10.96\\
2001-01-17T18:22:03&10.89& 132& 234&3.92&9.81\\
2001-02-17T03:14:09&6.36&  63& 153&4.51&10.53\\
2001-02-20T03:49:24&9.98& 273& 525&2.70&10.83\\
2001-02-20T23:41:18&6.58&  69& 150&7.01&11.05\\
2001-02-23T12:28:21&6.34& 168& 264&3.04&10.96\\
2001-03-01T14:42:45&6.10&  72& 123&6.41&9.52\\
2001-03-03T05:42:27&8.16& 171& 198&3.17&11.13\\
2001-03-04T06:09:15&5.73& 201& 312&2.11&10.27\\
2001-03-04T18:49:42&14.37&  78&  75&10.14&10.75\\
2001-03-05T16:40:21&14.48& 711& 516&1.13&10.73\\
2001-03-06T02:36:21&13.88& 735& 969&0.84&10.43\\
2001-03-11T08:13:09&14.53& 807&1077&0.91&10.84\\
2001-03-16T15:54:15&18.01& 297& 384&3.34&11.43\\
2001-03-18T04:26:09&5.81& 198& 333&2.51&9.68\\
2001-03-18T12:26:12&22.42& 264& 135&3.24&9.59\\
2001-04-13T08:30:06&5.86& 165& 354&2.05&10.70\\
2001-05-06T19:20:15&5.99& 126& 141&3.96&9.72\\
2001-05-16T09:44:03&15.96& 348& 270&1.89&11.64\\
2001-05-22T14:13:12&8.90& 414& 978&1.19&9.81\\
2001-05-28T14:10:27&7.32& 333& 333&1.74&11.05\\
2001-06-02T13:26:54&9.73& 138& 240&4.59&10.83\\
2001-06-05T08:01:06&9.68& 201& 225&3.06&11.45\\
2001-06-07T13:03:00&13.89& 189&  90&4.32&10.80\\
2001-06-10T17:26:06&11.51& 210& 267&2.08&9.93\\
2001-06-29T23:50:00&7.62& 126& 150&3.85&9.94\\
2001-07-06T06:01:33&3.68&  42&  36&4.38&9.71\\
2001-07-08T01:50:03&4.12& 153& 222&3.02&9.80\\
2001-07-13T07:50:12&7.62&  57& 180&9.83&11.04\\
2001-07-15T23:24:51&8.86& 174& 447&3.42&10.03\\
2001-07-23T01:44:18&8.28& 243& 162&2.42&10.00\\
2001-07-23T21:57:36&13.03& 102& 129&6.29&10.94\\
2001-07-29T19:10:42&13.07& 498& 426&1.91&10.48\\
2001-08-03T05:34:39&12.72& 345& 330&2.29&9.60\\
2001-08-05T19:54:00&17.32& 237& 336&4.10&11.06\\
2001-08-08T06:29:30&10.56&  63& 153&4.26&11.52\\
2001-08-09T04:38:18&20.86& 114&  78&6.60&11.53\\
2001-08-16T02:52:42&8.10&  78&  60&7.37&11.08\\
2001-08-16T17:46:36&8.98& 225& 405&2.61&10.91\\
2001-08-17T09:25:12&6.10&  54& 102&6.66&10.16\\
2001-09-02T04:08:48&14.16& 162& 114&3.53&11.19\\
2001-11-26T19:21:27&16.41& 141& 183&4.39&10.72\\
2002-03-26T15:21:30&31.13& 693&1524&1.42&10.22\\
2002-04-01T13:03:39&14.00& 180& 261&4.18&10.49\\
2002-04-02T16:28:36&12.34&  42&  93&14.49&11.22\\
2002-04-20T09:35:18&13.81& 738&1182&0.96&11.95\\
2002-04-24T10:53:15&21.55& 351& 441&3.07&10.62\\
2002-04-26T13:26:54&9.25& 159& 165&3.20&10.44\\
2002-05-02T03:32:33&17.14& 219&1089&4.33&10.90\\
2002-05-07T12:42:00&21.05& 171& 240&6.35&11.70\\
2002-05-12T09:54:12&38.16&1491& 969&0.80&10.47\\
2002-05-19T06:49:36&11.88& 138& 579&6.02&11.59\\
2002-05-25T05:47:00&14.01& 177& 159&3.98&11.25\\
2002-06-12T01:19:21&8.70& 144& 180&3.29&10.31\\
2002-06-14T00:02:51&9.55&  87& 228&5.55&11.20\\
2002-06-25T01:59:03&6.87&  54&  54&9.77&10.02\\
2002-06-26T17:18:45&4.53&  60& 201&6.09&10.58\\
2002-06-26T20:09:57&6.31& 135& 186&2.67&9.88\\
2002-06-29T12:53:21&8.52& 249& 246&2.08&10.35\\
2002-07-03T08:51:51&11.34& 129& 297&3.06&9.80\\
2002-07-11T02:39:57&16.24& 180& 117&4.96&11.69\\
2002-07-12T07:01:42&17.04& 894&1161&0.83&11.79\\
2002-07-12T19:45:27&8.22& 111& 195&4.18&10.91\\
2002-07-16T19:15:00&14.64& 141& 141&6.35&10.94\\
2002-09-24T20:03:36&17.83& 327& 267&3.46&11.68\\
2002-09-28T09:00:12&8.39& 132& 240&2.81&9.60\\
2002-09-28T10:01:39&7.86& 141& 273&4.05&11.20\\
2002-09-30T06:43:39&32.99& 468& 588&2.26&10.55\\
2002-10-06T02:26:33&11.98&  45&  60&12.41&11.47\\
2002-10-11T01:53:00&12.11& 420& 360&1.65&10.83\\
2002-11-05T05:54:06&13.77& 123&  54&6.16&11.69\\
2002-11-24T02:13:00&10.98& 213& 273&2.57&10.60\\
2002-11-25T22:33:00&10.62&  54& 120&5.80&11.06\\
2002-12-08T16:43:00&8.69& 171& 156&3.57&10.14\\
2002-12-27T16:52:45&6.55& 486& 477&1.06&9.37\\
\end{longtable}